\documentclass[journal]{IEEEtran}

\usepackage{amsmath,amssymb,amsfonts}

\usepackage{graphicx}
\usepackage{svg}
\usepackage{tikz}
\usetikzlibrary{arrows.meta,calc,positioning}
\usepackage{array}
\usepackage{multirow}
\usepackage{booktabs}
\usepackage{threeparttable}
\usepackage{wrapfig}
\usepackage{stfloats}
\usepackage{float}

\usepackage{xcolor}
\usepackage{listings}
\usepackage[most]{tcolorbox}

\usepackage[linesnumbered,ruled,vlined]{algorithm2e}
\usepackage{algpseudocode}

\usepackage{subcaption}

\usepackage{url}
\usepackage{textcomp}
\usepackage{pifont}
\usepackage{enumerate}
\usepackage{enumitem}
\usepackage{xr}
\usepackage{hyperref}

\usepackage{cite}

\usepackage{balance}

\def\eq {Equation~}

\definecolor{LightGray}{gray}{0.95}

\newtcolorbox[auto counter]{custombox}[1][]{
  enhanced,
  breakable,
  title={Listing \thetcbcounter},
  boxrule=0.5mm,
  left=0.5mm,
  right=0.5mm,
  top=0.5mm,
  bottom=0.5mm,
  #1
}

\newtcolorbox{summarybox}{
  enhanced,
  boxrule=0.3mm,
  left=0.5mm,
  right=0.5mm,
  top=0.5mm,
  bottom=0.5mm,
}

\newcommand{\mybox}[1]{%
  \begin{tcolorbox}[colback=white,colframe=black,lowerbox=invisible,savelowerto=\jobname_ex.tex, left=2pt, right=2pt, top=2pt, bottom=2pt]
    #1
  \end{tcolorbox}
}

\usepackage{listings}
\usepackage{xcolor}

\definecolor{LightGray}{gray}{0.95}

\lstdefinestyle{stacktrace}{
    backgroundcolor=\color{LightGray},
    frame=single,
    rulecolor=\color{black},
    basicstyle=\footnotesize\ttfamily,
    numbers=left,
    numberstyle=\tiny,
    stepnumber=1,
    numbersep=5pt,
    showstringspaces=false,
    breaklines=true,
    tabsize=2
}

\newcommand{\rev}[1]{\textcolor{black}{#1}}
\newcommand{\revtwo}[1]{\textcolor{black}{#1}}

\begin{document}

\title{Stack Trace-Based Crash Deduplication with Transformer Adaptation}

\author{
    Md Afif Al Mamun$^1$, Gias Uddin$^2$, Lan Xia$^3$, Longyu Zhang$^4$ \\ 
    $^1$ University of Calgary, $^2$ York University, $^{3,4}$ IBM Canada
}

\maketitle

\begin{abstract}
    Automated crash reporting systems generate large volumes of duplicate reports, overwhelming issue-tracking systems and increasing developer workload. Traditional stack trace-based deduplication methods—relying on string similarity, rule-based heuristics, or deep learning (DL) models—often fail to capture the contextual and structural relationships within stack traces. We propose \textbf{dedupT}, a transformer-based approach that models stack traces holistically rather than as isolated frames. dedupT first adapts a pretrained language model (PLM) to stack traces, then uses its embeddings to train a fully-connected network (FCN) to rank duplicate crashes effectively. Extensive experiments on real-world datasets show that dedupT outperforms existing DL and traditional methods (e.g., sequence alignment and information retrieval techniques) in both duplicate ranking and unique crash detection, significantly reducing manual triage effort. On four public datasets, dedupT improves Mean Reciprocal Rank (MRR) often by over 15\% compared to the best DL baseline and up to 10\% over traditional methods while achieving higher Receiver Operating Characteristic Area Under the Curve (ROC-AUC) in detecting unique crash reports. Our work advances the integration of modern natural language processing (NLP) techniques into software engineering, providing an effective solution for stack trace-based crash deduplication. 
    \textbf{Replication Package.} \url{https://github.com/afifaniks/dedupt}
\end{abstract}

\section{Introduction}
\label{sec:introduction}

Software issues are generally reported through (1) human-submitted reports and (2) automated crash reports. Human-reported issues typically include textual descriptions detailing the issue, expected and observed behavior, and may include attachments such as images or videos. In contrast, automated crash reports are generated by crash reporting tools (e.g., Sentry\footnote{\url{https://sentry.io/}}, CrashPad\footnote{\url{https://chromium.googlesource.com/crashpad/crashpad/}}) embedded within software or large-scale test pipelines, which log failures directly into an issue-tracking system (ITS). However, these automated systems often overwhelm ITS platforms by generating numerous duplicate crash reports for the same issue, requiring developers to manually review and triage them, which is a time-consuming process. For instance, Mozilla Firefox received 2.2 million issues in the first week of 2016, the majority being duplicates \cite{campbell2016unreasonable}, while 72\% of crash reports in the IntelliJ Platform were found to be duplicates \cite{rodrigues2022tracesim}. In such scenarios, grouping similar crashes together is essential, a process known as \textit{crash deduplication}.

Unlike human-written reports with detailed descriptions, automated crash reports primarily contain technical data like stack traces and crash dumps. In such cases, stack traces serve as the key source for identifying duplicate issues \cite{Sabor2017DURFEXAF, Brodie2005MLMMWCS05, Dhaliwal2011, rodrigues2022tracesim, Khvorov2021S3MSS}. A stack trace \( S \) is a recorded snapshot of the sequence of function or method calls that were active at the time an error or exception occurred. Formally, a stack trace can be represented as an ordered list of \textit{frames} or \textit{stack frames}  $S = [f_1, f_2, ..., f_n]$, where each frame \( f_i \) corresponds to a function or method call in the execution stack. Each frame \( f_i \) typically contains information like the function/method name, file location, and line number where the call was made. Figures \ref{fig:java-stack-trace} and \ref{fig:cpp-stack-trace} include two examples of stack traces for Java and C++, respectively.
\begin{figure}[t]
    \centering
    \begin{lstlisting}[style=stacktrace, language=Java]
Exception in thread "main" java.lang.NullPointerException
    at com.example.MyClass.myMethod(MyClass.java:10)
    at com.example.MyClass.main(MyClass.java:5)
    \end{lstlisting}
    \vspace{-1em}
    \caption{Example of a Java stack trace.}
    \label{fig:java-stack-trace}
    \vspace{-1em}
\end{figure}

\begin{figure}[t]
    \centering
    \begin{lstlisting}[style=stacktrace, language=C]
#0  0x40990b in crash_func() at example.c:6
#1  0x40250a in inter_function() at example.c:10
#2  0x40150a in main() at example.c:15
    \end{lstlisting}
    \vspace{-1em}
    \caption{Example of C++ stack trace.}
    \label{fig:cpp-stack-trace}
    \vspace{-1em}
\end{figure}

Stack traces provide debugging information by capturing the call hierarchy leading to a failure \cite{schroter2010dostacktraces, wong2014boosting}. While numerous stack-trace-based crash deduplication (STCD) methods exist, accuracy remains a challenge. Traditional approaches primarily compare frames through lexical or sequence-matching signals \cite{rodrigues2022tracesim, Sabor2017DURFEXAF, Brodie2005MLMMWCS05, dang2012rebucket}, while deep learning (DL) approaches such as S3M \cite{Khvorov2021S3MSS} and DeepCrash \cite{deepcrash} first learn representations of individual frames or their subframes and then compose them into a stack-trace representation using recurrent networks.

\begin{figure*}[t]
    \centering
    \resizebox{0.8\textwidth}{!}{%
%
\begin{tikzpicture}[
  x=1cm, y=1cm, >=Latex,
  font=\sffamily\scriptsize,
  code/.style={font=\fontsize{6.3}{7.6}\selectfont\ttfamily},
  cell/.style={code, draw, rounded corners=1.5pt, line width=.5pt, align=center,
               inner xsep=2pt, inner ysep=2pt},
  same/.style={cell, draw=teal!60!black, fill=teal!8},
  diff/.style={cell, draw=orange!78!black, fill=orange!18, line width=.8pt},
  absent/.style={cell, draw=black!35, fill=none, text=black!45,
                 dash pattern=on 1pt off 1.4pt,
                 font=\fontsize{6.3}{7.6}\selectfont\sffamily\itshape},
  generic/.style={cell, draw=black!55, fill=black!5, text=black!68, line width=.7pt},
  exc/.style={cell, draw=black!72, fill=black!9,
              font=\fontsize{6.3}{7.6}\selectfont\ttfamily\bfseries},
  rid/.style={font=\sffamily\scriptsize\bfseries, text=orange!62!black},
  tag/.style={font=\fontsize{5.8}{6.8}\selectfont\sffamily,
              inner xsep=2.5pt, inner ysep=.6pt, fill=white},
  badge/.style={font=\fontsize{5.8}{6.8}\selectfont\sffamily\bfseries,
                text=white, fill=black!55,
                rounded corners=1.5pt, inner xsep=3pt, inner ysep=1.6pt},
  region/.style={rounded corners=2.5pt, line width=.5pt, draw=orange!70!black,
                 fill=orange!4, dash pattern=on 2.5pt off 1.8pt},
  verdict/.style={font=\sffamily\scriptsize\bfseries},
]

\begin{scope}[shift={(0,0.82)}]
  \draw[rounded corners=3pt, line width=.85pt, draw=red!55!black, fill=white]
    (0,0) rectangle (8.45,6.05);
  \node[anchor=west, font=\sffamily\footnotesize\bfseries, text=red!55!black] at (.28,5.75)
    {(a) Same exception and first frame; different call context};

  \node[exc, text width=7.749cm, minimum height=4.4mm] at (4.225,5.20)
    {BasicIndexOutOfBoundsException};
  \node[tag, anchor=west, text=black!62] at (.56,5.42) {same exception};

  \node[same, text width=7.749cm, minimum height=4.4mm] at (4.225,4.53)
    {BasicEList.get};
  \node[tag, anchor=west, text=teal!45!black] at (.56,4.75)
    {same first frame (failure site)};

  \draw[region] (.28,2.26) rectangle (8.17,4.01);
  \node[tag, anchor=west, text=orange!62!black] at (.56,4.01)
    {different call context immediately below the first frame};

  \node[rid] at (2.2675,3.75) {Report \#107736};
  \node[rid] at (6.1825,3.75) {Report \#114094};

  \node[diff, text width=3.434cm, minimum height=4.6mm] at (2.2675,3.26)
    {AdaptorEditor.createPages};
  \node[diff, text width=3.434cm, minimum height=4.6mm] at (6.1825,3.26)
    {EcoreEditor.createModel};

  \node[absent, text width=3.434cm, minimum height=4.6mm] at (2.2675,2.67)
    {no counterpart frame};
  \node[diff, text width=3.434cm, minimum height=4.6mm] at (6.1825,2.67)
    {EcoreEditor.createPages};

  \draw[rounded corners=2pt, line width=.7pt, draw=black!55, fill=black!5]
    (.28,.80) rectangle (8.17,2.06);
  \node[badge, anchor=north west] at (.40,1.98)
    {27 identical consecutive framework frames -- limited discriminative evidence};
  \node[code, text=black!68, align=center] at (4.225,1.26)
    {MultiPageEditorPart.createPartControl\\[-1pt]
     \ldots\\[-1pt]
     WorkbenchPage.openEditor};

  \node[verdict, text=red!55!black] (va) at (4.42,.42)
    {NON-DUPLICATE (different buckets)};
  \coordinate (ia) at ([xshift=-3.4mm]va.west);
  \draw[red!55!black, line width=.7pt] (ia) circle (1.7mm);
  \draw[red!55!black, line width=1pt, line cap=round]
    ($(ia)+(-.85mm,-.85mm)$) -- ($(ia)+(.85mm,.85mm)$)
    ($(ia)+(-.85mm,.85mm)$) -- ($(ia)+(.85mm,-.85mm)$);
\end{scope}

\draw[-{Latex[length=2.6mm,width=1.9mm]}, draw=black!55, line width=1pt]
  (8.90,5.62) -- (8.90,1.62);
\node[rotate=-90, font=\sffamily\scriptsize, text=black!62, fill=white,
      inner xsep=2pt, inner ysep=1.5pt] at (8.90,3.62) {call depth};

\begin{scope}[shift={(9.35,0.82)}]
  \draw[rounded corners=3pt, line width=.85pt, draw=green!42!black, fill=white]
    (0,0) rectangle (8.45,6.05);
  \node[anchor=west, font=\sffamily\footnotesize\bfseries, text=green!42!black] at (.28,5.75)
    {(b) Matching call context with one inserted frame};

  \node[exc, text width=7.749cm, minimum height=4.4mm] at (4.225,5.20)
    {BundleException};
  \node[tag, anchor=west, text=black!62] at (.56,5.42) {same exception};

  \node[same, text width=7.749cm, minimum height=10mm] at (4.225,4.32)
    {BundleContextImpl.startActivator\\[-1pt]
     \ldots\\[-1pt]
     SecureAction.start};
  \node[tag, anchor=west, text=teal!45!black] at (.56,4.82)
    {identical prefix};

  \draw[region] (.28,2.26) rectangle (8.17,3.56);
  \node[tag, anchor=west, text=orange!62!black] at (.56,3.56)
    {one inserted frame inside an otherwise matching path};

  \node[rid] at (2.2675,3.30) {Report \#287720};
  \node[rid] at (6.1825,3.30) {Report \#423034};

  \node[absent, text width=3.1cm, minimum height=5.5mm] at (2.2675,2.71)
    {no additional frame};
  \node[diff, text width=3.7cm, minimum height=5.5mm] at (6,2.71)
    {BundleLoader.setLazyTrigger};

  \node[same, text width=7.749cm, minimum height=12.6mm] at (4.225,1.43)
    {EclipseLazyStarter.postFindLocalClass\\[-1pt]
     \ldots\\[-1pt]
     ClassLoader.loadClass};
  \node[tag, anchor=west, text=teal!45!black] at (.56,2.06)
    {10 identical consecutive frames after the insertion};

  \node[verdict, text=green!42!black] (vb) at (4.42,.42)
    {DUPLICATE (Bucket \#287720)};
  \coordinate (ib) at ([xshift=-3.4mm]vb.west);
  \draw[green!42!black, line width=.7pt] (ib) circle (1.7mm);
  \draw[green!42!black, line width=1pt, line cap=round, line join=round]
    ($(ib)+(-.9mm,.05mm)$) -- ($(ib)+(-.25mm,-.75mm)$) -- ($(ib)+(.95mm,.85mm)$);
\end{scope}

\node[same, minimum width=4.2mm, minimum height=2.4mm, inner sep=0pt]
  at (.79,.36) {};
\node[anchor=west, text=black!72] at (1.03,.36) {identical frames};

\node[diff, minimum width=4.2mm, minimum height=2.4mm, inner sep=0pt]
  at (3.71,.36) {};
\node[anchor=west, text=black!72] at (3.95,.36) {frames that differ};

\node[generic, minimum width=4.2mm, minimum height=2.4mm, inner sep=0pt]
  at (6.87,.36) {};
\node[anchor=west, text=black!72] at (7.11,.36)
  {identical generic framework frames};

\node[absent, minimum width=4.2mm, minimum height=2.4mm, inner sep=0pt]
  at (12.00,.36) {};
\node[anchor=west, text=black!72] at (12.24,.36) {no counterpart frame};

\node[code, text=black!50] at (15.35,.36) {\ldots};
\node[anchor=west, text=black!72] at (15.55,.36) {omitted frames};

\end{tikzpicture}%
    }
    \caption{\revtwo{Examples of call-context ambiguity from the Eclipse dataset. Panel (a) shows non-duplicate reports with the same exception and first frame but different immediate call contexts; panel (b) shows duplicate reports with one inserted intermediate frame. Class and package qualifiers and intermediate frames are omitted for readability.}}
    \label{fig:call-context-motivation}
\end{figure*}

\rev{\revtwo{Representing stack traces effectively presents two complementary challenges. First, an individual frame can be ambiguous without its surrounding calls. Figure \ref{fig:call-context-motivation}(a) illustrates two reports from the Eclipse crash report dataset \cite{rodrigues2022tracesim}, where Reports \#107736 and \#114094 belong to different ground-truth \textit{buckets}, i.e., groups of reports triaged as manifestations of the same issue, despite throwing the same exception, beginning with the same frame, and subsequently sharing 27 consecutive framework frames. Their different editor-specific calls immediately after the first frame distinguish the two call contexts. Conversely, Figure \ref{fig:call-context-motivation}(b) shows that Reports \#287720 and \#423034 belong to the same ground-truth bucket even though the latter contains an additional intermediate call within an otherwise closely matching path. Thus, interpreting the ordered relationships among frames can distinguish different failure contexts without requiring exact frame-by-frame equality. Second, frames are not necessarily equally informative. Schr\"oter et al. found that, among bugs whose fixes modified a method appearing in the submitted stack trace, approximately 90\% involved a method within the top ten frames \cite{schroter2010dostacktraces}. As Figure \ref{fig:call-context-motivation}(a) further illustrates, reports in different ground-truth buckets may share long sequences of framework calls, which provide limited evidence for distinguishing their failure contexts. These observations suggest that a useful stack-trace representation should preserve relationships among multiple ordered frames while emphasizing the failure-relevant portion of the trace, rather than treating frames independently or assigning equal importance to the entire trace.}}

Recent advancements in transformer architecture \cite{vaswani2017attention} have revolutionized contextual analysis in textual contents. Pretrained language models (PLMs) \cite{devlin-etal-2019-bert, roberta2019, matthew2018elmo, Deberta} have achieved state-of-the-art (SOTA) performance across a variety of contextual NLP tasks, including sentiment analysis, text summarization, and semantic similarity \cite{bashiri2024comprehensive, Zhang2019BERTScoreET, li2024pre}. Among these, SentenceBERT (SBERT) \cite{Reimers2019SentenceBERTSE} stands out due to its ability to generate high-quality sentence or passage-level embeddings for textual-similarity analysis while being exponentially faster than the original BERT \cite{devlin-etal-2019-bert}. To the best of our knowledge, there is a lack of exploration into how these models can be specifically adapted to understand nuances in stack traces for crash deduplication. 

\revtwo{We propose \textbf{dedupT} (\textbf{\underline{dedup}}lication with \textbf{\underline{T}}ransformer), a stack-trace-based approach that ranks the existing duplicate buckets most likely to contain a duplicate of an incoming crash report. dedupT retains an empirically selected number of top frames in their relative order and encodes them jointly as one stack-trace representation. It adapts a transformer encoder using duplicate and non-duplicate stack trace pairs, aggregates multiple stack traces into report-level representations, and uses a learned classifier to produce a duplication score between the query and each historical report. The maximum report-level score within each bucket determines the final bucket ranking. The scope of dedupT is therefore the retrieval of duplicate crash reports and ranking duplicate buckets. We evaluate this approach on four widely used datasets against ten established baselines.}

Overall, dedupT outperformed all baseline methods, including sequence matching, information retrieval (IR), and deep learning (DL) approaches, in Mean Reciprocal Rank (MRR) and various Recall Rate@k metrics, demonstrating the effectiveness of the transformer-based approach for STCD. Specifically, it achieved MRR gains of up to 10\% over the strongest IR baseline and over 15\% on three of the four datasets relative to the strongest DL baseline. Additionally, dedupT proved significantly more effective in identifying unique crashes, achieving an ROC-AUC improvement of over 20\% compared to most alignment-based methods, while remaining competitive with other DL models. 

\noindent\textbf{\rev{Contributions.}}
\rev{This paper makes the following contributions:}

\begin{itemize}
    \item \rev{We propose \textbf{dedupT}, a transformer-based approach for stack trace–based crash deduplication that models stack traces holistically rather than as isolated frames.}
    
    \item \rev{We adapt a pretrained embedding model to the stack-trace domain using contrastive fine-tuning, improving the discriminative quality of stack trace embeddings.}
    
    \item \rev{We design a parametric aggregation strategy to combine multiple stack traces within a crash report, capturing both the most relevant trace pair and the overall contextual information.}

\end{itemize}

\rev{The remainder of this paper is organized as follows. Section \ref{sec:related-work} reviews related work on stack trace–based crash deduplication. Section \ref{sec:motivation} motivates the key design choices, and Section \ref{sec:approach} describes their implementation. Section \ref{sec:evaluation-setup} defines the experimental setup and evaluation metrics. Section \ref{sec:results} presents the evaluation results. Section \ref{sec:discussion} \rev{presents a manual study of how dedupT's rankings support duplicate triage and} discusses comparisons with LLM-based approaches, retrieval time, and applicability to other languages. \revtwo{Section \ref{sec:threats} represents the threats to the validity of the study.} Finally, Section \ref{sec:conclusions} concludes the paper and outlines future work.}


\section{Related Work}
\label{sec:related-work}

\noindent\textbf{Sequence matching and information retrieval based techniques.} Stack trace-based crash deduplication has been fairly studied in the literature. Early approaches mostly address this as a string matching problem. Brodie et al. \cite{Brodie2005MLMMWCS05} proposed Needleman-Wunsch based algorithm to compare stack traces. Modani et al. \cite{Modani2007} explored different methods for comparing stack traces including prefix matching, edit distance, etc., along with an indexing strategy for efficient retrieval. Bartz et al. \cite{Bartz2008FindingSF} used logistic regression, and Dhaliwal et al. \cite{Dhaliwal2011} introduced a two-step approach: generating stack trace signatures followed by Levenshtein distance computation to enhance similarity detection. These methods primarily rely on lexical matching, which may struggle with structurally different yet semantically similar traces.

Many approaches also adopted TF-IDF to find similar crash reports \cite{Lerch2013, campbell2016unreasonable, rodrigues2022tracesim}. Among these, Tracesim \cite{vasiliev2020tracesim} is the first to combine TF-IDF, Levenshtein distance, and machine learning (ML) to optimize hyperparameters. The same authors later proposed another version of Tracesim \cite{rodrigues2022tracesim} that provides a pipeline to evaluate a lot of existing approaches, including \cite{dang2012rebucket, Modani2007, Brodie2005MLMMWCS05, Sabor2017DURFEXAF}. This version of Tracesim also handled scenarios when multiple stack traces exist in a crash report. Later, Rodrigues et al. proposed FaST \cite{rodrigues2022fast}, which works in linear time by aligning stack frames closer to the top, considering these are the most important frames. While FaST achieves higher throughput, its retrieval accuracy is similar to TraceSim.

\noindent\textbf{Deep learning-based approaches.} Deep learning (DL) has been comparatively underexplored in stack trace-based crash deduplication. One of the earliest DL approaches, S3M \cite{Khvorov2021S3MSS}, uses a siamese biLSTM to encode each tokenized stack-trace sequence and a fully connected network trained with RankNet to predict pairwise similarity. DeepCrash \cite{deepcrash} learns subframe-based frame representations, encodes their sequence with a biLSTM using deep metric learning, and clusters the resulting vectors using a tuned similarity threshold. For Linux kernel crash classification, Abaci-finder \cite{shi2022abaci} introduces kstack2vec to represent function-name semantics together with kernel-specific offset information, followed by an attention-based biLSTM that considers frame context and key frames. More recently, Shibaev et al. \cite{shibaev2024stack} train a BPE tokenizer and hierarchical biLSTMs to encode tokens into frame vectors and frames into stack-trace embeddings for approximate-nearest-neighbor retrieval. Their second-stage biLSTM reranker marks frames that occur identically in a pair of traces. KERMIT \cite{yang2025kermit} instead applies kernel-specific call-trace filtering and full-parameter fine-tuning of BERT to adapt its semantic representations to Linux kernel crash deduplication.

\revtwo{Existing DL approaches have limitations that can affect duplicate retrieval. S3M maps complete frames through a predefined dictionary, while DeepCrash learns a subframe vocabulary; unseen units therefore become unknown tokens, which DeepCrash reports can limit performance \cite{Khvorov2021S3MSS,deepcrash}. Shibaev et al. and dedupT mitigate this problem through subword tokenization \cite{shibaev2024stack,kudo2018sentencepiece}, however Shibaev et al. train their BPE vocabulary and recurrent encoder on each dataset, whereas dedupT adapts a general-purpose pretrained encoder. Moreover, these approaches first vectorize frames or subframes and then compose their sequence using biLSTMs \cite{Khvorov2021S3MSS,deepcrash,shi2022abaci,shibaev2024stack}. This models frame order, but not token-level interactions across frame boundaries; Shibaev et al.'s reranker introduces pairwise information through exact-frame repetition indicators only after frame encoding. dedupT instead supplies the selected ordered frames jointly to a pretrained transformer and contrastively adapts the resulting stack-trace embedding. DeepCrash also requires a validation-tuned clustering threshold and leaves multi-trace aggregation as future work \cite{deepcrash}; the cutoff determines cluster assignments, while additional traces may be ignored. dedupT directly ranks buckets using learned pairwise scores and parametrically aggregates multiple stack traces in a crash report. Finally, Abaci-finder explicitly incorporates kernel-specific function-offset information, while BERT-based KERMIT relies on kernel-specific call-trace filtering and full-model tuning on kernel reports \cite{shi2022abaci,yang2025kermit}. However, these pipelines assume kernel trace fields and noise patterns that differ from user-space Java and C/C++ traces. dedupT instead adapts a general-purpose transformer representation strategy across Java and C/C++ crash datasets.}

\revtwo{\textit{In summary, dedupT combines supervised contrastive adaptation of general-purpose pretrained transformers with joint contextualization of the selected ordered frames and learned report-level aggregation of multiple traces. Subword tokenization, including strategies such as SentencePiece \cite{kudo2018sentencepiece}, additionally mitigates OOV effects for previously unseen identifiers.}}
\section{Key Design Choices in dedupT}
\revtwo{This section motivates two decisions that distinguish dedupT's representation from directly matching stack frames: (1) encoding the selected frames jointly with a transformer to preserve context across their ordered sequence and (2) adapting the embedding space to stack-trace duplicate relationships. Section~\ref{sec:approach} then describes how these decisions are implemented in the complete ranking pipeline.}
\label{sec:motivation}

\subsection{\rev{Use of Transformers}}
\begin{figure}[htbp]
    \centering
    \includegraphics[width=0.75\linewidth]{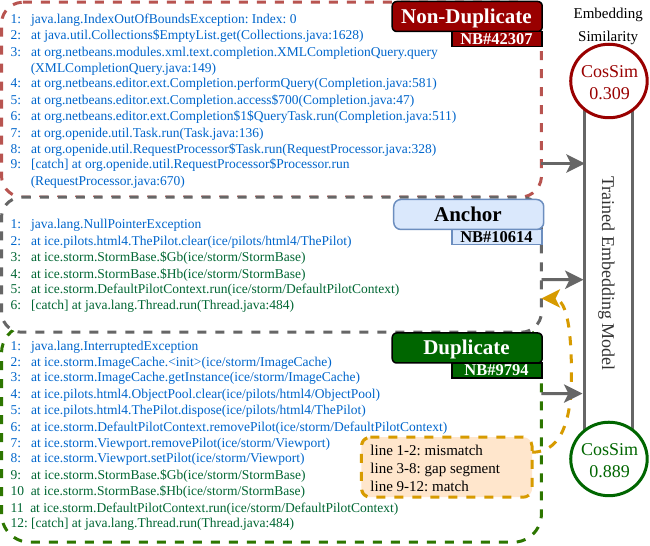}
    \caption{Cosine similarity of stack trace embeddings between a duplicate and a non-duplicate pair.}
    \label{fig:example-pair}
\end{figure}
Duplicate crash report detection is essential for grouping similar reports and aiding bug triaging in large-scale software projects. Traditional approaches to this problem primarily rely on information retrieval techniques (such as TF-IDF) \cite{Lerch2013, Sabor2017DURFEXAF, vasiliev2020tracesim, rodrigues2022fast, campbell2016unreasonable, dang2012rebucket} and sequence-matching algorithms (such as edit distance and global alignment) \cite{Brodie2005MLMMWCS05, Modani2007, Bartz2008FindingSF, Dhaliwal2011}. However, these methods struggle with structural variations in stack traces, particularly when only a few frames are shared between duplicate reports or when the mismatches are penalized as gaps \cite{rodrigues2022tracesim}. Furthermore, they often require extensive manual feature engineering hyperparameter tuning, making them time-consuming for large-scale deployment. For example, in our evaluation of TraceSim \cite{rodrigues2022tracesim}, optimizing the hyperparameters for a studied dataset (Netbeans) took over 12 hours.

Figure \ref{fig:example-pair} illustrates an example from the Netbeans dataset, where a stack trace from Report ID \href{https://bz.apache.org/netbeans/show_bug.cgi?id=10614}{\#10614} has a known duplicate (Report ID \href{https://bz.apache.org/netbeans/show_bug.cgi?id=9794}{\#9794}) and a non-duplicate stack trace (Report ID \href{https://bz.apache.org/netbeans/show_bug.cgi?id=42307}{\#42307}). \rev{One challenge in duplicate detection using sequence matching is evident in this example--the duplicate pair shares only four stack frames (green-colored, lines 9–12 in the duplicate and 3–6 in the anchor), while the rest of the frames, including exceptions (blue-colored), are different. According to the Tracesim \cite{rodrigues2022tracesim} alignment algorithm, lines 1–2 between duplicate and the anchor are treated as \textit{mismatches}, lines 3–8 form a contiguous \textit{gap} segment inserted to align the shared subroutines (9-12), and only lines 9–12 are \textit{matched}, which would lead to a low similarity score due to multiple gap penalties.} Similarly, methods that rely on the direct alignment of stack frames \cite{rodrigues2022tracesim} or keyword matching \cite{Brodie2005MLMMWCS05} would assign a low similarity score to this pair, failing to recognize their underlying relationship. Deep learning (DL) models are also explored to capture the sequential relationships for STCD. Specifically, LSTM networks have been explored \cite{Khvorov2021S3MSS, deepcrash} to encode stack frames. However, LSTMs generally struggle with longer sequences in stack traces due to their limited memory capacity \cite{karita2019comparative}. Stack traces often contain deeply nested call structures with functionally equivalent but lexically different paths, making it difficult for LSTMs to generalize across variations. For instance, two stack traces may exhibit different function names (e.g., \textit{ClassA.getUser()} vs. \textit{ClassB.findUser()}) or variations in framework-level calls while originating from the same root cause. Such differences arise due to alternative execution paths, middleware abstractions, or reordered stack frames. Since LSTMs process sequences token by token, they are highly sensitive to lexical variations and lack an inherent understanding of structural similarities, making them less effective in deduplication.


Building on these findings, we evaluate transformer-based embedding models that use self-attention \cite{vaswani2017attention} to capture structural and semantic patterns in stack traces for duplicate detection. Specifically, we adapt a pre-trained SBERT model \cite{Reimers2019SentenceBERTSE} to generate high-quality stack trace embeddings. Unlike traditional DL methods (e.g., LSTMs), SBERT captures contextual relationships, identifying duplicates despite lexical differences. To refine its ability to distinguish similar and dissimilar traces (i.e., adaptation), we fine-tune it using a contrastive learning \cite{chen2020contrastive} objective. As shown in Figure \ref{fig:example-pair}, this adapted model assigns a high similarity score (0.889) to a duplicate pair while assigning a very low score to an unrelated trace (0.309). \rev{While we do not use embedding similarity directly to identify duplicates, rather we train a downstream classifier on top of the embedding model}, this analysis highlights how transformers can generalize beyond exact frame matching and produce high-quality embeddings that capture execution-flow patterns, enabling downstream accurate ranking and grouping of duplicate crash reports.

\subsection{Stack trace adaptation of embedding models}
\begin{figure}[htbp]
    \centering
    \includegraphics[width=0.65\linewidth]{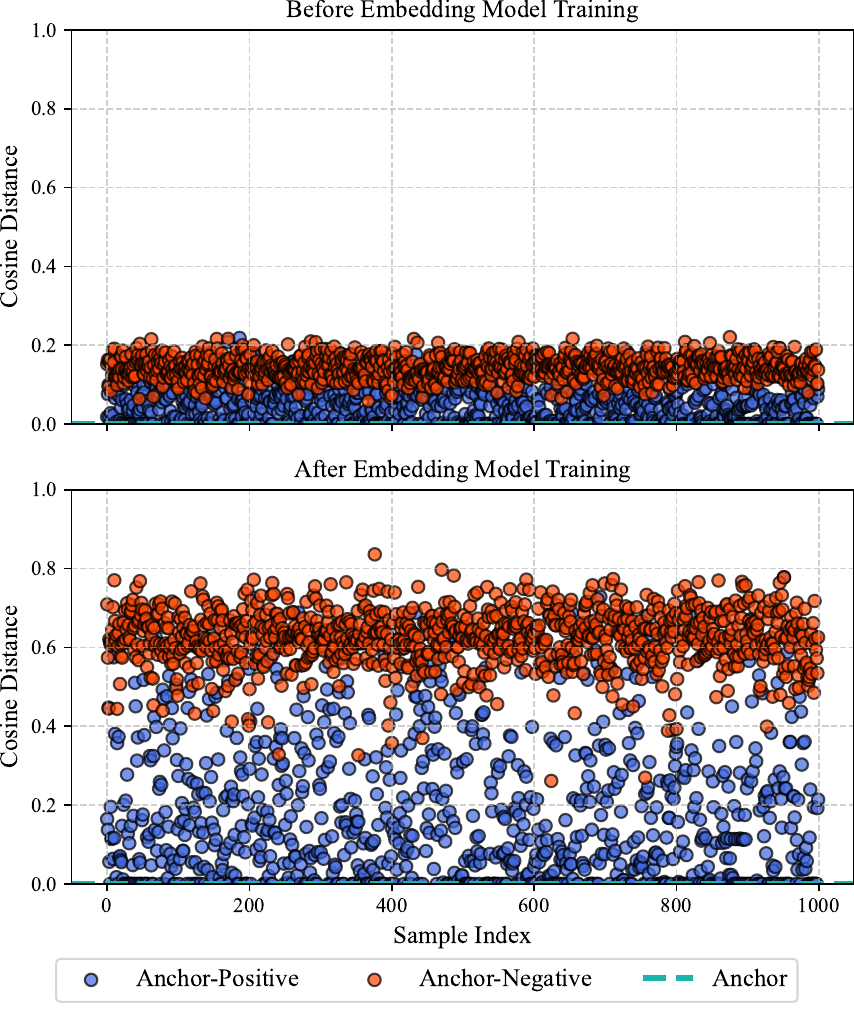}
    \caption{Cosine distance (\(1-\mathrm{CosSim}\)) between stack-trace embeddings before and after embedding-model adaptation on NetBeans dataset.}
    \label{fig:embedding-distance}
    \vspace{-1em}
\end{figure}
PLM-based embedding models (EMs) are typically trained on natural language (NL) corpora like MSMARCO \cite{bajaj2018msmarcohumangenerated} and WikiHow \cite{Fader14wikihow}, etc. However, stack traces follow a completely different structure with hierarchical syntax of method calls, class names, and file paths. General-purpose EMs may fail to capture fine-grained similarities in stack traces, which are important for tasks like duplicate crash detection. Small differences in file paths, method signatures, or exception types can signal distinct issues, while semantically identical stack traces may vary due to runtime factors. 

Thus, separately fine-tuning or adapting an EM on stack traces enables the model to learn domain-specific representations that are better suited for crash deduplication. Unlike generic PLMs, which may fail to capture the structural and semantic nuances of stack traces, a fine-tuned model can differentiate better between meaningful variations in execution paths, exception types, and method calls, etc.

We utilize contrastive learning to improve the model's ability to distinguish between duplicate and non-duplicate stack traces. Contrastive learning optimizes the embedding space by pulling semantically similar stack traces closer together while pushing dissimilar ones apart. This ensures that stack traces belonging to the same bucket are mapped closer together while unrelated ones are separated. As illustrated in Figure \ref{fig:embedding-distance}, the embedding distance between negative pairs (i.e., stack trace pairs that are not duplicates) increases significantly after fine-tuning the model on a studied dataset, indicating that the model learns a more discriminative embedding space, reducing false positives in duplicate detection. Additionally, when we fine-tune an EM separately, it is possible to precompute embeddings for all stack traces and cache them for both faster training and retrieval.
\section{Methodology}
\label{sec:approach}
\begin{figure*}[htbp]
    \centering
    \includegraphics[width=0.7\linewidth]{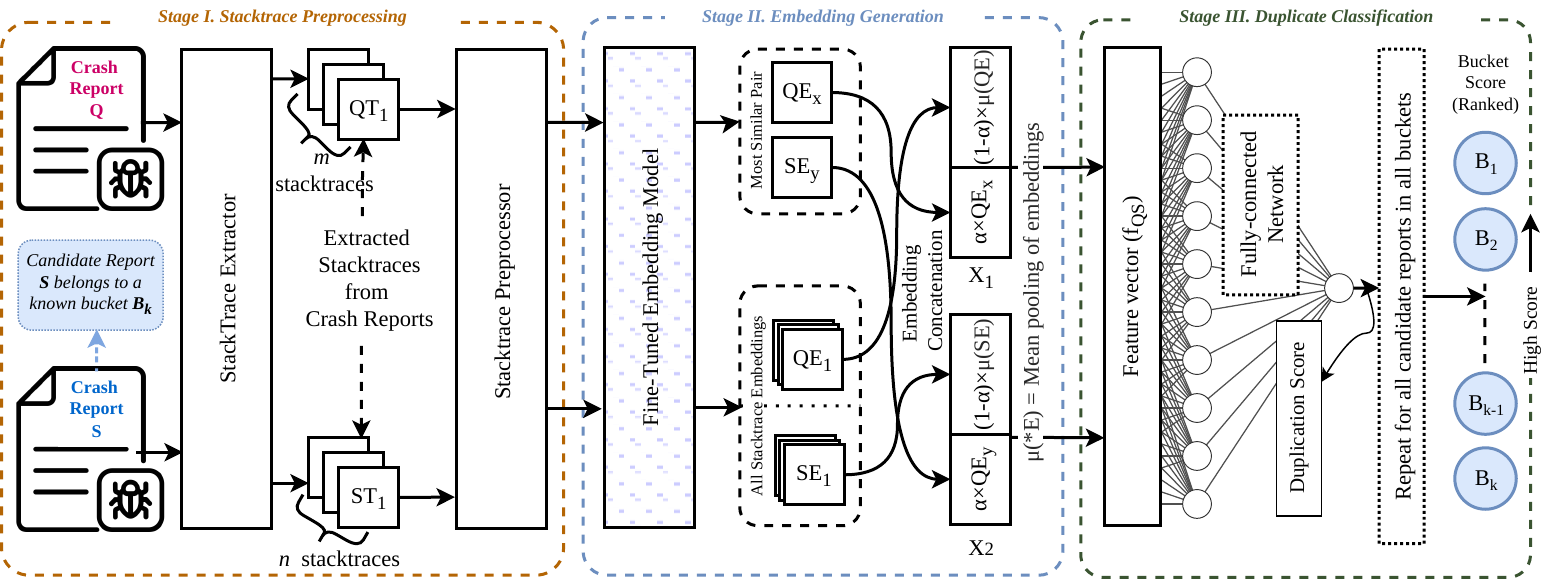}
    \caption{Architecture diagram of dedupT. $QT_{x}, QE_{x}$ represents x-th stack trace and its embedding from report Q.}
    \label{fig:main-arch}
\end{figure*}

\rev{\textbf{dedupT} performs stack-trace-based crash deduplication (STCD) through three main stages (Figure \ref{fig:main-arch}). Stage I (Stacktrace Preprocessing) extracts and preprocesses stack traces from crash reports. Stage II (Embedding Generation) encodes each stack trace using a fine-tuned embedding model and aggregates multiple traces into a single report-level representation. Stage III (Duplicate Classification) computes pairwise duplication scores between the query report and candidate reports using a fully connected network, then performs bucket-level ranking by using scores within each duplicate bucket.}

\rev{While dedupT adopts standard preprocessing components (e.g., stack-trace extraction and trimming) from established methods, it introduces three methodological advances in how stack traces are represented and aggregated for crash deduplication.} \rev{\revtwo{First, dedupT introduces a task-specific adaptation strategy that fine-tunes a pretrained transformer-based embedding model on stack traces using supervised contrastive learning (Section \ref{sec:embedding-model-adaptation}). To the best of our knowledge, dedupT is the first crash-deduplication approach to adapt a general-purpose pretrained embedding model to stack traces using supervised contrastive learning over duplicate and non-duplicate pairs, learning an embedding space in which traces from the same bucket are positioned closer together while traces from different buckets are moved farther apart. This distinguishes dedupT from methods that rely on off-the-shelf embedding models or lexical/alignment-based similarity, as well as BERT-based approaches optimized for direct duplicate classification \cite{yang2025kermit}. Second, dedupT supplies the selected ordered frames directly to a pretrained transformer to produce a single contextualized stack-trace embedding (Section \ref{sec:embedding-model-generation}). In comparison, existing DL approaches \cite{Khvorov2021S3MSS,deepcrash,shibaev2024stack} first learn representations of individual frames or subframes and then compose their sequence using biLSTMs.} Third, we propose a \emph{parametric multi-stack aggregation} mechanism for reports containing multiple stack traces (Section \ref{sec:multi-report-representation}). Instead of heuristic pooling, we learn aggregation weights to emphasize the most informative traces.}

\subsection{\rev{Stage I. Stacktrace Preprocessing}}
\label{sec:stack-preprocessor}

\subsubsection{Stacktrace Extractor}
\rev{Stack traces differ across programming languages. A typical Java stack trace (Figure \ref{fig:java-stack-trace}) begins with an exception message, followed by a sequence of stack frames that represent method calls, each containing details such as the class name, method name, and line number. In contrast, C++ stack traces—often obtained using tools like \texttt{backtrace} or debuggers such as \texttt{gdb}—usually provide memory addresses alongside function names (Figure \ref{fig:cpp-stack-trace}). Following Rodrigues et al. \cite{rodrigues2022tracesim}, we extract stack traces from crash reports using the Java stacktrace extractor from Lerch et al. \cite{Lerch2013} and \href{https://metacpan.org/pod/Parse::StackTrace}{Parse::StackTrace} for C++.}

\subsubsection{Stacktrace Preprocessor}
\label{sec:stack-trace-processor}
Since pretrained models impose a fixed input sequence length, long stack traces must be pruned to avoid exceeding this limit. Redundant frames consume valuable token space without adding meaningful information, which can degrade embedding quality. To address this, we design a four-step preprocessing pipeline, where only the \textit{Frame Cleaning} stage differs between different languages to account for their structural differences.

\noindent\textbf{Duplicate Removal.} We remove duplicate frames in a stack trace following the approach of Brodie et al. \cite{Brodie2005MLMMWCS05}, which eliminates consecutive frames with the same subroutine name under the same package or module. This is particularly helpful in cases of recursion and repeated library calls. 

\begin{figure}[htbp]
    \centering
    \includegraphics[width=0.9\linewidth]{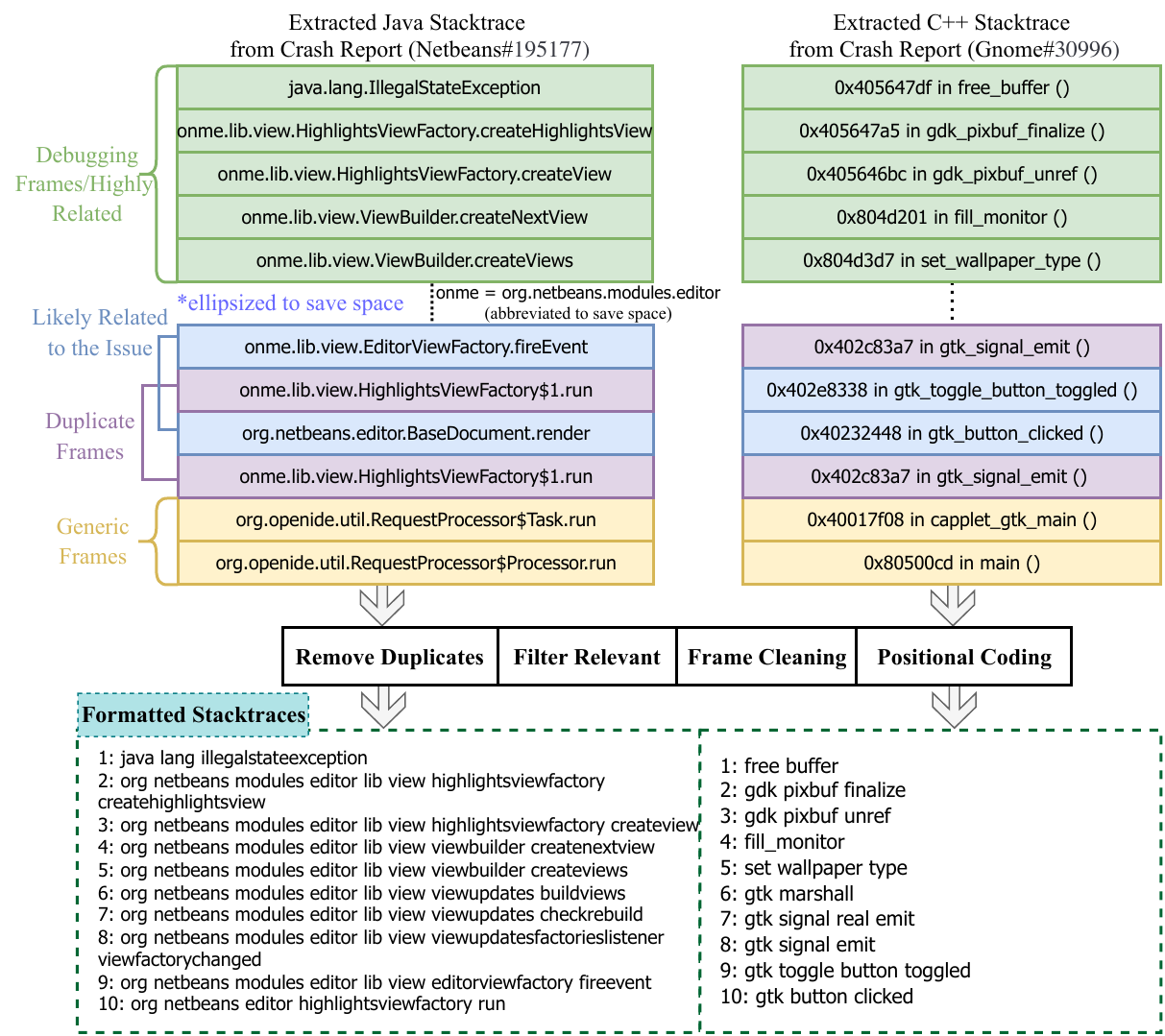}
    \caption{\revtwo{Stacktrace preprocessor applied to illustrative stack traces synthesized from NetBeans Report \#195177 and Gnome Report \#30996. The displayed traces are truncated for readability; ellipses omit intermediate frames that remain present in the full original traces provided in the replication package.}}
    \label{fig:stack-processor}
\end{figure}

\noindent\textbf{Sampling Relevant Frames.} \revtwo{Prior work found that fix-relevant methods are concentrated near the top of stack traces, with about 90\% occurring within the top ten frames among bugs fixed in a method listed in the trace \cite{schroter2010dostacktraces}.} Frames from standard libraries or deep recursion may also add little discriminative information. Moreover, pretrained language models impose token length constraints, making it impractical to include full traces when stacktraces get longer. \rev{The number of frames considered for embedding generation directly impacts the quality of the resulting embeddings. For example, retaining too few frames may omit important contextual information related to the root cause, while retaining too many may introduce noise and reduce discriminative power. To balance informativeness with efficiency, we retain only the top $N$ frames and treat $N$ as a tunable parameter. The optimal value of $N$ is determined empirically through ablation analysis (see Section \ref{sec:ablation}).}


\noindent\textbf{Frame Cleaning.} PLMs are primarily trained on natural language, so they may misinterpret stack-trace syntax. In Java traces, for example, dots (.) indicate package hierarchies rather than sentence boundaries, and in C++ traces, special characters or debugger-specific prefixes (e.g., \_\_IA\_\_, \_\_GI\_\_) are common. If left unprocessed, these symbols can be misinterpreted as sentence breaks or irrelevant tokens. To mitigate this, we normalize stack frames by replacing dots and special characters with whitespace, lowercasing all tokens, and treating each frame as a sentence while considering the entire trace as a passage. Following prior work \cite{Sabor2017DURFEXAF, Khvorov2021S3MSS}, we also evaluate different levels of trimming for Java stack traces and discuss their impact in Section~\ref{sec:ablation}.



\noindent\textbf{Positional Coding.} Finally, each stack frame is assigned positional information to encode execution order, allowing the model to distinguish between frames and capture the sequential structure of function calls essential for recognizing stack trace patterns. 

An example illustrating how stack traces are cleaned and preprocessed for both Java and C++ is shown in Figure~\ref{fig:stack-processor}. While this work focuses on two popular statically typed languages—Java and C++, the approach should be easily extendable to other languages by adapting the frame cleaning stage to handle the syntax and conventions.

\subsection{Stage II. Embedding Generation}
\label{sec:embedding-model}

\rev{Stage II aims to generate semantic embeddings for stack traces and aggregate them into report-level representations. This stage consists of two parts: (i) a one-time \textit{offline training} phase that adapts a pretrained embedding model to the stack-trace domain, and (ii) an embedding generation phase that uses the trained model to produce embeddings during inference and classifier training.}

\subsubsection{\rev{Offline Embedding Model Adaptation}}
\label{sec:embedding-model-adaptation}
\rev{The embedding model adaptation procedure is performed offline exclusively without the classifier, prior to any inference, and produces a fixed encoder that is reused for all subsequent embedding generation.}

\noindent\textbf{Training Data Construction.} We employ supervised fine-tuning to adapt a pretrained embedding model to the stack-trace domain using stack-trace pairs derived from human-labeled crash datasets \cite{rodrigues2022tracesim}. Let \(\mathcal{B} = \{B_1, B_2, \dots, B_k\}\) be the set of all crash buckets, where each bucket \(B_i\) contains stack traces corresponding to duplicate crash reports.  

For each training sample, we define an \textit{anchor} \(x_a \in B_i\) as a reference stack trace. The data generation process involves constructing an equal number of positive and negative pairs with respect to the anchor:  

\begin{itemize}  
    \item \textbf{Positive pairs} (\(x_a, x_p\)): Both stack traces belong to the same bucket \(B_i\), i.e., \(x_a, x_p \in B_i\), indicating duplicate crashes.  
    \item \textbf{Negative pairs} (\(x_a, x_n\)): The stack traces belong to different buckets, i.e., \(x_a \in B_i\) and \(x_n \in B_j\) when \(i \neq j\), indicating non-duplicate crashes. We construct negative pairs by sampling from a random subset of stack traces drawn from the top 50 buckets identified by Lerch and Mezini \cite{Lerch2013}, excluding the correct pair.
\end{itemize}  

The objective of the consecutive fine-tuning is to optimize the embedding model such that the anchor \(x_a\) is mapped closer to its positive counterpart \(x_p\) than to its negative counterpart \(x_n\) in the embedding space. \rev{Notably, each training instance corresponds to a contiguous sequence of top frames (e.g., top 10) of a stack trace, rather than an individual frame in isolation. This is particularly important for contrastive learning as a single frame stripped of its call context loses its discriminative power, as the same generic frame (e.g., \texttt{Thread.run} or \texttt{java.lang.IllegalArgumentException}) can appear across crashes of entirely different root causes. Using such frames as independent training instances would introduce false positives and negatives within the same batch, producing contradictory gradient signals that directly undermine the contrastive objective.}



\noindent\textbf{Contrastive Fine-Tuning.} The embedding model is fine-tuned/adapted using the positive and negative pairs generated in the earlier stage through a contrastive objective, Multiple Negatives Ranking Loss (MNR) \cite{henderson2017efficient}, defined as:
\begin{equation}
\begin{aligned}
\mathcal{L}_{\text{MNR}}
&= -\frac{1}{N}\sum_{i=1}^{N}
\log
\frac{
\exp\left(\text{CosSim}(f_\theta(x_a^i),f_\theta(x_p^i))/\tau\right)
}{
\sum_{x\in\mathcal{C}_i}
\exp\left(\text{CosSim}(f_\theta(x_a^i),f_\theta(x))/\tau\right)
},
\end{aligned}
\end{equation}
where \(N\) is the batch size, \(\mathcal{C}_i\) is the candidate set containing the positive \(x_p^i\) and the negatives \(x_n\) for anchor \(x_a^i\), and \(\tau\) is a temperature scaling factor that controls the sharpness of the cosine similarity scores. \rev{This loss function encourages the encoder model \(f_\theta\) to maximize the similarity between the anchor (\(x_a\)) and the positive (\(x_p\)) while minimizing the cosine similarity between the anchor and the negatives (\(x_n\)).} This whole adaptation process uses only data from the training split.

\subsubsection{\rev{Stack-Trace Embedding Generation}}
\label{sec:embedding-model-generation}
Once the embedding model is fine-tuned with stack traces, it can be used to extract meaningful features from them. A significant advantage of separate fine-tuning is that all precomputed stack trace embeddings can be cached or stored, enabling efficient retrieval and eliminating the need to repeatedly calculate embeddings during training. Since embedding generation is the most expensive part of the pipeline, this approach allows for faster classifier training while maintaining scalability.  

However, a crash report \(Q\) may contain multiple stack traces (MST) \(\{s_1^Q, s_2^Q, \dots, s_n^Q\}\), as noted in \cite{rodrigues2022tracesim}. Similarly, a candidate report \(S\) may have multiple stack traces \(\{s_1^S, s_2^S, \dots, s_m^S\}\). To compare the similarity of the reports \(Q\) and \(S\), it is essential to consider the relationships between all their respective stack traces.  

\subsubsection{\rev{Report-Level Representation with MST}}
\label{sec:multi-report-representation}
A simple and efficient way to combine them is by concatenating all stack traces before embedding generation:  $S_{\text{concat}} = \{s_1 \oplus s_2 \oplus \dots \oplus s_n\}$. However, this method is constrained by the embedding model’s maximum sequence length and is more preferable when frames are short (e.g., C++ subroutines). Since Java stack traces are often verbose and require a large number of tokens, this approach becomes impractical for Java-based crash reports. In such cases, another approach is to aggregate all stack trace embeddings using \textit{mean pooling}, defined as:  
\begin{equation}
    \mu(QE) = \frac{1}{n} \sum_{i=1}^n f(s_i^Q), \quad \mu(SE) = \frac{1}{m} \sum_{j=1}^m f(s_j^S),
\end{equation}
where \(f(s)\) is the embedding of stack trace \(s\), and \(\mu(QE)\), \(\mu(SE)\) represent the aggregated embeddings for \(Q\) and \(S\), respectively. However, mean pooling assumes equal contribution from all stack traces, ignoring their varying relevance and averaging over noise, which may dilute meaningful representations.

\noindent\textbf{MST Aggregation.}  
To address these challenges, we propose a hybrid approach called Parametric Max-Mean that captures both the most relevant stack trace pairs (max pooling) and the context of all stack traces (mean pooling). First, we identify the most similar pair of stack traces, \((s_i^Q, s_j^S)\), with the fine-tuned embedding model using cosine similarity and construct features \(f(s_i^Q)\) and \(f(s_j^S)\) from their embedding.
\begin{equation}
\label{eq:max-sim}
    (s_i^Q,s_j^S)=\arg\max_{i,j}\text{CosSim}(f(s_i^Q),f(s_j^S)).
\end{equation}
Then, we compute the mean-pooled embeddings of all stack traces, \(\mu(QE)\) and \(\mu(SE)\), which aggregate the embeddings of all stack traces from crash reports \(Q\) and \(S\), respectively. To allow flexibility in determining the importance of these features, we introduce a trainable priority parameter \(\alpha \in [0, 1]\). The final representation of stack traces for crash report \(Q\) is computed by concatenating both:
\begin{equation}
\label{eq:parametric-max-mean}
    f_Q = \alpha \cdot f(s_i^Q) \parallel (1 - \alpha) \cdot \mu(QE)
\end{equation}


This approach ensures that the model captures both the most relevant stack trace information and the broader context of all stack traces while allowing the importance of each component to be learned during training. 

\subsection{\rev{Stage III. Duplicate Classification}}
\rev{In this stage, we determine whether an incoming crash report corresponds to an existing duplicate group (i.e., bucket). While the model computes duplication scores between pairs of reports, the final scoring is done at the bucket level, where each bucket represents a set of reports that have been previously labeled as duplicates of one another.}

\noindent\textbf{Classifier Architecture.} The classifier is implemented as a fully connected network (FCN) designed to take feature vectors derived from two crash report stack traces and output a duplication score. Following previous approaches \cite{Homma2017DetectingDQ, Khvorov2021S3MSS}, we construct the feature vector $f_{QS}$ for the classifier using the following equation:  
\begin{equation}
    f_{QS} = \big( |f_Q - f_S|, \frac{f_Q + f_S}{2}, f_Q \odot f_S \big),
\end{equation}
where \(f_Q \text{ and } f_S\) are the representations of stack traces computed in \eq\ref{eq:parametric-max-mean}. The rationale behind this feature construction is to capture a comprehensive representation of both the differences and similarities between the two embeddings. These features are then passed through an FCN defined in Equation \ref{eq:sim-score} to calculate the duplication score.

\begin{equation}
\label{eq:sim-score}
    \text{DupScore} = \Phi_2\left(\sigma\left( \Phi_1 \left( \text{Dropout}(f_{QS}) \right) \right)\right)
\end{equation}

Where $\Phi_i$ is linear layer $i$ and $\sigma$ is an activation function. The classifier includes a dropout \cite{srivastava2014dropout} layer with a probability \(p = 0.1\) to prevent overfitting, followed by two linear layers. The first linear layer reduces the feature dimension by half, followed by a ReLU activation, and the second layer outputs a scalar duplication score for the report pair. This score is calculated for all candidate reports from existing buckets and the query report to rank the candidates.

\noindent\textbf{Bucket-Level Scoring.}  
\rev{Although the model produces a pairwise duplication score \(DupScore(Q, S)\) between two reports \(Q\) and \(S\), crash deduplication is typically performed at the bucket level \cite{Khvorov2021S3MSS, rodrigues2022tracesim, rodrigues2022fast}. In issue-tracking systems, reports are organized into buckets, where each bucket \(B_i\) represents a group of reports that have been identified as duplicates of one another. Therefore, the practical goal of deduplication is not merely to assess pairwise similarity, but also to determine whether an incoming report belongs to one of these existing duplicate groups.} \rev{Given a query report \(Q\), we compute \(DupScore(Q, s)\) for all historical reports \(s \in B_i\). For each bucket \(B_i \in \mathcal{B}\), we define its bucket-level score as:}

\begin{equation}
    \label{eq:bucket-score}
    BucketScore(Q, B_i) = \max_{s \in B_i} DupScore(Q, s).
\end{equation}

\rev{\revtwo{The max operator reflects the intuition that if \(Q\) receives a high $DupScore$ to any report within \(B_i\), then \(Q\) likely belongs to that duplicate group.} We compute \(BucketScore(Q, B_i)\) for all existing buckets and sort them in descending order based on this score.}



\noindent\textbf{Classifier Training.} The classifier is trained using RankNet \cite{burges2005learning} loss. We generate triplets (\(x_a, x_p, x_n\)), where \(x_a\) is the anchor, \(x_p\) is a randomly selected stack trace from the same bucket (i.e., duplicate), and \(x_n\) is a non-duplicate stack trace selected following the process described in Section \ref{sec:embedding-model}. Specifically, for each triplet, we compute the following RankNet loss:

\begin{equation}
\mathcal{L}
=
\log\left(
1+\exp\left(
\text{DupScore}(x_a,x_n)
-
\text{DupScore}(x_a,x_p)
\right)
\right).
\end{equation}

\revtwo{where \(\text{DupScore}(x_a,x_p)\) and \(\text{DupScore}(x_a,x_n)\) are the predicted duplication scores for the anchor-positive and anchor-negative pairs, respectively. The loss minimizes ranking inconsistency by encouraging the positive sample to receive a higher duplication score than the negative sample.}

The model is trained with a batch size of 25 with Adam optimizer \cite{Kingma2014AdamAM} with a learning rate of $10^{-4}$ to minimize the RankNet loss. All the training and evaluation were done on Intel(R) Xeon(R) Gold 6148 CPU@2.40GHz with an NVIDIA V100 16GB GPU, although the VRAM usage typically remained $< 3\text{GB}$ during training.



\section{Evaluation of DedupT}
\label{sec:evaluation-setup}

\subsection{Research Questions} 

We answer the following research questions (RQs):
\begin{enumerate}[leftmargin=30pt, label=\textbf{RQ\arabic{*}.}]
    \item \textbf{Duplicate Retrieval Performance:} How does dedupT perform in duplicate crash retrieval compared to existing baselines?
    \item \textbf{Detection of Unique Crashes:} How effectively does dedupT identify and distinguish unique crash reports?
    \item \textbf{Embedding Model Adaptation:} How much does adaptation improve performance, and how does the adapted model compare with state-of-the-art (SOTA) embedding models?
    \item \textbf{Ablation Study:} How do different design choices impact the overall performance?
\end{enumerate}

\subsection{Baselines}
We adopt the evaluation pipeline from Rodrigues et al. \cite{rodrigues2022tracesim}, which includes popular baselines: Rebucket \cite{dang2012rebucket}, Tracesim \cite{rodrigues2022tracesim}, DURFEX \cite{Sabor2017DURFEXAF}, Moroo \cite{moroo2017reranking}, PrefixMatching \cite{Modani2007}, Brodie \cite{Brodie2005MLMMWCS05}, Needleman-Wunsch (NW), and TF-IDF. For DL baselines, we include S3M \cite{Khvorov2021S3MSS}, which uses an LSTM-based bi-encoder for frame representations, and DeepCrash (DC) \cite{deepcrash}, which tokenizes frames into subframes and learns stack trace representations via LSTMs. Since DC was originally designed for clustering, we modified it for ranking to enable comparison. Additionally, as DURFEX is specific to Java stack traces, we exclude GNOME and Ubuntu from its evaluation.

\noindent\textbf{Excluded Methods.} Our evaluation focuses strictly on stack trace-based crash deduplication, excluding methods that incorporate additional information \cite{he2020duplicate, xiao2020hindbr, rodrigues2020soft, sun2011rep}, as our target scenario assumes the absence of natural language content. Some baselines were omitted due to code availability, complexity, or time constraints. Specifically, we exclude FaST \cite{rodrigues2022fast} as it prioritizes retrieval speed (which is out of scope for this study), where it performs similarly to TraceSim. We also could not include DeepLSH \cite{remil2024deeplsh} and Shibaev et al. \cite{shibaev2024stack} due to reproducibility issues with their published code.

\subsection{Datasets}


\begin{table}[h]
    \centering
    \caption{Dataset statistics.}
    \resizebox{8.5cm}{!}{\begin{tabular}{lccccccc}
        \hline
        \textbf{Dataset} & \textbf{Period} & \textbf{\# Dup} & \textbf{\# Report} & \textbf{\# Bucket} &\textbf{\% MST} \\
        \hline
        Eclipse & 2001/10/11 - 2018/12/31  & 8,332 & 55,968 & 47,636 &  24.37\%\\
        Netbeans & 1998/09/25 - 2016/12/31 & 13,703 & 65,417 & 51,714 & 61.37\% \\
        Gnome & 1998/01/02 - 2011/12/31 & 117,216 & 218,160 & 100,944 & 80.14\%\\
        Ubuntu & 2007/05/25 - 2015/10/18 & 11,468 & 15,293 & 3,825 & 0.03\%\\
        \hline
    \end{tabular}
    }
    \label{table:dataset_stats}
\end{table}

We use large-scale crash report datasets from Campbell et al. \cite{campbell2016unreasonable} and Rodrigues et al. \cite{rodrigues2022tracesim}, derived from the issue-tracking systems of Netbeans, Eclipse, Gnome, and Ubuntu. Netbeans and Eclipse datasets contain Java stack traces, while Gnome and Ubuntu consist of C/C++ traces. These reports were manually categorized into unique issues or duplicates based on human triage decisions. Table \ref{table:dataset_stats} summarizes the dataset statistics, including the number of reports, duplicates, buckets, and percentage of multiple stack traces (MST) in crash reports. Following DL-baseline S3M \cite{Khvorov2021S3MSS}, we used (4200, 700, 140) days for training, testing, and validation on the Netbeans, Eclipse, and Gnome datasets and (2350, 650, 50) days for Ubuntu since it has data for only $<3100$ days. \rev{To ensure fair comparison, all baseline methods were trained and evaluated using the same temporal train/validation/test partitions described above. For instance, on the Netbeans dataset, we reproduced TraceSim’s MRR as 0.700, which closely matches the 0.704 reported in its original \href{https://github.com/irving-muller/TraceSim_EMSE/blob/main/test_results.ipynb}{replication package}, confirming consistency with prior results. We also provide detailed instructions in our replication package for generating data splits and running the baselines.}

\begin{figure}[t]
    \centering
    \includegraphics[width=0.8\linewidth]{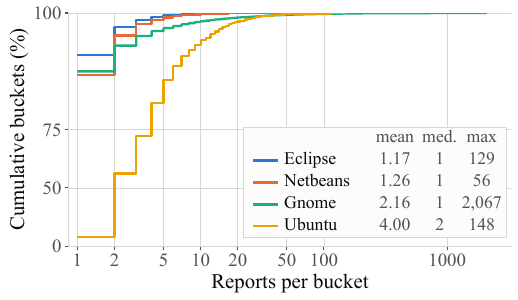}
    \caption{\revtwo{Cumulative distribution of bucket sizes.}}
    \label{fig:bucket-size-dist}
    \vspace{-1em}
\end{figure}

\revtwo{Since crash deduplication ranks buckets rather than individual reports for each query, we also characterize how many reports these buckets hold. Figure~\ref{fig:bucket-size-dist} shows the distribution of bucket sizes in each dataset. The datasets fall into two regimes. In Eclipse, Netbeans, and Gnome the median bucket holds a single report and between 86.8\% and 91.0\% of buckets are singletons, since most reports are never marked as a duplicate of anything. Ubuntu differs sharply, with only 7.7\% singletons and a median of two reports. Every dataset also has a long tail, the largest bucket holding 129 reports in Eclipse and 2{,}067 in Gnome, but the tail is thin. The proportion of buckets containing more than ten reports ranges from 0.1\% in Eclipse to 5.9\% in Ubuntu, and buckets of more than a hundred reports are 0.2\% of Gnome and rarer elsewhere. These distributions bound the number of reports a retrieved bucket can contain, but not the number a developer must actually read, since a bucket groups reports that are deemed as manifestations of identical issues \cite{dang2012rebucket}.}

\subsection{Embedding Model Evaluation}
\noindent\textbf{Embedding Model Selection.} Choosing an embedding model is critical, as performance largely depends on the training corpora, with domain-specific models typically outperforming general-purpose ones \cite{asudani2023impact}. Since no embedding model exists for stack traces, we adapt a PLM to this domain. Section~\ref{sec:ablation} shows that our fine-tuning approach achieves strong performance regardless of the underlying PLM. We select a few candidate models from the HuggingFace repository that can be used with the Sentence Transformer library. These models are: all-mpnet-v2 (mpnet) \cite{huggingface-all-mpnet}, all-distilroberta-v1 (distilroberta) \cite{huggingface-distilroberta}, bge-base-en (bge) \cite{xiao2024cpack}, etc. However, for most experiments of this paper, we selected \texttt{bge} as our preferred embedding model due to its slightly better performance on our evaluation.

\noindent\textbf{Embedding Model Evaluation Metrics.} \rev{While the ultimate objective of this work is crash deduplication at the bucket level, dedupT relies on learning discriminative stack-trace embeddings that serve as the foundation for downstream duplicate classification and ranking. The quality of the embedding space directly affects how effectively the supervised classifier can separate duplicate and non-duplicate pairs and how reliably duplicate buckets are ranked.}

Therefore, before integrating embeddings into the full classification and ranking pipeline, we assess whether the learned representations meaningfully distinguish positive (duplicate) and negative (non-duplicate) stack-trace pairs. Specifically, we measure the correlation between predicted cosine similarity scores and ground-truth labels (1 for duplicate, 0 for non-duplicate) using Pearson and Spearman correlation coefficients. Higher Pearson values indicate stronger linear alignment with the stack trace pair labels (i.e., positive/negative), while higher Spearman values capture better monotonic ordering of pairs by similarity. We additionally compute Euclidean similarity to assess whether duplicate stack traces cluster more closely in the embedding space. \revtwo{However, this is only a diagnostic analysis of embedding quality; final duplicate ranking uses the learned \(DupScore\) (Equation \ref{eq:sim-score}) and \(BucketScore\) (Equation \ref{eq:bucket-score}).}

\subsection{Evaluation Metrics for Deduplication}
\noindent\textbf{Duplicate Ranking.} Following previous studies \cite{Khvorov2021S3MSS, rodrigues2022tracesim, rodrigues2022fast}, we employed Mean Reciprocal Rank (MRR) and Recall Rate (RR@k) to evaluate the retrieval ranking of our approach compared to the baselines. MRR is defined as the average of the reciprocal ranks of the first relevant result:

\begin{equation}
    MRR = \frac{1}{|Q|} \sum_{i=1}^{|Q|} \frac{1}{rank_i}
\end{equation}

where \( Q \) is the set of queries and \rev{\(rank_i\) is the rank position of the ground-truth bucket of query \(i\) in the predicted bucket ranking.}

Recall Rate at rank \( k \) (RR@k) \rev{measures the proportion of queries for which the ground-truth bucket appears within the top \(k\) ranked buckets}:

\begin{equation}
    RR@k = \frac{1}{|Q|} \sum_{i=1}^{|Q|} \mathbb{1} (rank_i \leq k)
\end{equation}

where \( \mathbb{1}(\cdot) \) is an indicator function that equals 1 if the condition holds and 0 otherwise. We set \( k \leq 10 \) since, in practical scenarios, users rarely consider recommendations beyond the top 10 results. \rev{Finally, we report statistical significance using the Wilcoxon signed-rank test. We compare our approach against the strongest baseline (i.e., the baseline achieving the highest MRR on the corresponding dataset). The test is performed on the per-query reciprocal rank scores, enabling paired comparisons across identical queries. We consider results statistically significant at $p < 0.01$.}

\begin{table*}
    \small
    \caption{Performance comparison of different approaches on benchmark datasets. ($^{*}$) indicates a statistically significant difference between dedupT and the strongest baseline (highest MRR), based on a paired Wilcoxon signed-rank test over per-query reciprocal ranks (\(p < 0.01\)).}
    \centering
    \resizebox{18cm}{!}{
    \begin{tabular}{lcccc|cccc|cccc|cccc}
        \toprule
        \multirow{2}{*}{\textbf{Approach}} & \multicolumn{4}{c|}{\textbf{Netbeans}} & \multicolumn{4}{c|}{\textbf{Eclipse}} &
        \multicolumn{4}{c|}{\textbf{Gnome}}& 
        \multicolumn{4}{c}{\textbf{Ubuntu}}\\
        \cmidrule(lr){2-5} \cmidrule(lr){6-9} \cmidrule(lr){10-13}\cmidrule(lr){14-17}
        & \textbf{MRR} & \textbf{RR@1} & \textbf{RR@5} & \textbf{RR@10} 
        & \textbf{MRR} & \textbf{RR@1} & \textbf{RR@5} & \textbf{RR@10}
        & \textbf{MRR} & \textbf{RR@1} & \textbf{RR@5} & \textbf{RR@10}
        & \textbf{MRR} & \textbf{RR@1} & \textbf{RR@5} & \textbf{RR@10}\\
        \midrule
        Tracesim & 0.700 & 0.623 & 0.809 & 0.856 & 0.753 & 0.685 & 0.833 & 0.857 & 0.710 & 0.643 & 0.790 & 0.823 & 0.736 &	 0.686 &	 0.801 &	 0.818\\
        Rebucket & 0.692 & 0.622 & 0.774 & 0.800 & 0.734 & 0.673 & 0.808 & 0.833 & 0.551 & 0.500 & 0.612 & 0.641 & 0.620 &	 0.568 &	 0.676 &	 0.740 \\
        PrefixMatching & 0.698 & 0.618 & 0.786 & 0.824 & 0.722 & 0.668 & 0.790 & 0.820 & 0.624 & 0.586 & 0.665 & 0.689 & 0.679 &	 0.625 &	 0.750 &	 0.780 \\
        Needleman & 0.646 & 0.554 & 0.755 & 0.808 & 0.722 & 0.657 & 0.804 & 0.833 & 0.691 & 0.628 & 0.740 & 0.765 & 0.675 &	0.625 &	0.723 &	0.770 \\
        Durfex & 0.622 & 0.525 & 0.740 & 0.790 & 0.682 & 0.600 & 0.781 & 0.823 & -- & -- & -- & -- & -- & -- & -- & -- \\
        Moroo & 0.689 & 0.609 & 0.787 & 0.820 & 0.728 & 0.664 & 0.803 & 0.825 & 0.664 & 0.596 & 0.743 & 0.778 & 0.708 &	 0.645 &	 0.781 &	 0.817 \\ 
        Brodie & 0.619 & 0.521 & 0.735 & 0.792 & 0.707 & 0.655 & 0.769 & 0.788 & 0.649	& 0.614	& 0.688	& 0.706 & 0.719 &	0.672 &	0.770 &	0.797\\
        TF-IDF & 0.604 & 0.510 & 0.717 & 0.765 & 0.651 & 0.568 & 0.749 & 0.787 & 0.687 & 0.611 & 0.733 & 0.809 & 0.693 &	 0.612 &	 0.792 &	 0.830 \\
        S3M & 0.618	& 0.530 & 0.724	& 0.766 & 0.706	& 0.640 & 0.770 & 0.799 & 0.565 & 0.499 & 0.630 & 0.673 & 0.675 & 0.610 & 0.751 & 0.787\\
        DeepCrash & 0.658 & 0.567 & 0.768 & 0.816 & 0.732 & 0.670 & 0.805 & 0.827 & 0.413 & 0.365 & 0.446 & 0.491 & 0.395 & 0.335 & 0.448 & 0.505 \\
        \midrule
        \textbf{dedupT} & \textbf{0.771$^{*}$} & \textbf{0.681} & \textbf{0.878} & \textbf{0.918} 
        & \textbf{0.791$^{*}$} & \textbf{0.720} & \textbf{0.880} & \textbf{0.902} & \textbf{0.742$^{*}$} & \textbf{0.676} & \textbf{0.808} & \textbf{0.841} & \textbf{0.786$^{*}$} & \textbf{0.744} & \textbf{0.838} & \textbf{0.866}\\
        \bottomrule
    \end{tabular}
    }
    \label{tab:rr-comparison}
    \vspace{-1em}
\end{table*}

\noindent\textbf{Unique Crash Detection.} To evaluate how well the models perform in identifying unique crash reports (i.e., not a duplicate of any existing bucket), we adopted the Receiver Operating Characteristic - Area Under the Curve (ROC-AUC) metric, which quantifies the model's ability to distinguish between duplicate and unique reports. The ROC-AUC score is computed as:

\begin{equation}
    \text{ROC-AUC} = \int_{0}^{1} TPR(FPR) \, dFPR
\end{equation}

where the True Positive Rate (TPR) and False Positive Rate (FPR) are defined as:

\begin{equation}
    TPR = \frac{TP}{TP + FN}, \quad FPR = \frac{FP}{FP + TN}
\end{equation}

Where \( TP \), \( FP \), \( TN \), and \( FN \) denote true positives, false positives, true negatives, and false negatives, respectively. The ROC curve plots TPR vs. FPR across thresholds, while AUC quantifies the likelihood of ranking a relevant report above a non-relevant one. 

\section{Results}
\label{sec:results}
\subsection{\textbf{RQ1: Duplicate Retrieval Performance}}
To evaluate dedupT for duplicate ranking, we compare its MRR and RR@k against baseline methods. As shown in Table II, dedupT consistently outperforms existing approaches across these metrics. Its overall ranking improvement over the strongest baseline (highest MRR) on each dataset is statistically significant, as confirmed by the Wilcoxon signed-rank test over paired per-query reciprocal ranks (\(p < 0.01\)). For instance, it achieves a 10\% higher MRR than the closest baseline, Tracesim, on the Netbeans dataset.

Despite being DL-based, S3M and DeepCrash underperform compared to other methods. DeepCrash performs better than S3M on Java datasets but struggles with C/C++ due to subframe tokenization, which aligns well with Java stack frames but lacks structure for C++ coroutines and modular information. S3M suffers from its reliance on a single stack trace even when multiple traces are available, along with a tokenization strategy prone to OOV issues with unseen stack frames.

Traditional methods show varying effectiveness but often demand extensive hyperparameter tuning or manual feature engineering, limiting their practicality in real-world deployment.


\noindent\textbf{Sensitivity to Data Shifts.} Rakha et al. \cite{rakha2018} noted that deduplication performance varies with data distribution shifts. To systematically study the impact of data selection on model performance, we employed three different chronological splits on two stack trace datasets: Netbeans (Java) and Ubuntu (C++). These datasets vary significantly in size and type, which allows us to assess how data distribution shift may impact transformer based models. In our approach, we train and test (both embedding model and duplicate classifier) on different time-based subsets, gradually increasing the training data to observe its effect on performance. For Netbeans, we extend the training period by 350 days ($\approx 1$ year) at each step, while for Ubuntu, we adjust it by 100 days, given its total span of only $\approx$ 3,100 days. 
\begin{table}
\centering
\small
\caption{Performance on different time-splits across datasets.}
\resizebox{7cm}{!}{
\begin{tabular}{lccccc}
\toprule
\textbf{Dataset} & \textbf{Train Days} & \textbf{Val Days} & \textbf{Test Days} & \textbf{MRR} & \textbf{RR@1} \\
\midrule
\multirow{3}{*}{Netbeans} & 3850 & 140 & 700 & 0.763 & 0.673 \\
                            & 4200 & 140 & 700 & 0.771 & 0.681 \\
                           & 4550 & 140 & 700 & 0.777 & 0.690 \\
                           
\hline
                           \multicolumn{4}{r}{\textbf{Mean $\mu$}} & 0.770 & 0.681 \\
                           \multicolumn{4}{r}{\textbf{SD $\sigma$}} &       0.005 &   0.006     \\
\midrule
\multirow{3}{*}{Ubuntu}    & 2250 & 50 & 700 & 0.761 & 0.706 \\
                           & 2350 & 50 & 650 & 0.786 & 0.744 \\
                           & 2450 & 50 &  600 & 0.838 & 0.790 \\
\midrule
                           \multicolumn{4}{r}{\textbf{Mean $\mu$}} &      0.795  &      0.746  \\
                           \multicolumn{4}{r}{\textbf{SD $\sigma$}} &     0.032   &   0.034   \\
\bottomrule
\end{tabular}
}
\label{tab:time-split-performance}
\vspace{-1em}
\end{table}

The results summarized in Table \ref{tab:time-split-performance} show two key trends: (1) The model exhibits minimal variance across different training splits on the Netbeans dataset, with a low standard deviation (SD) for both MRR and RR@1. This stability suggests that the model effectively generalizes due to the larger volume of data, and (2) the model shows a higher variance in performance on the Ubuntu dataset. As the training period increases, both MRR and RR@1 improve notably, indicating that the model benefits from additional training data.

\mybox{\textbf{RQ1:} \small{dedupT outperforms all baseline methods with higher MRR and RR@k, while showing stable performance across different data splits on larger datasets.}} 

\subsection{\textbf{RQ2: Detection of Unique Reports}}
Accurate detection of unique reports is vital for efficient triaging. Misclassifying them as duplicates can delay fixes, while correct identification ensures proper attention. Following Rodrigues et al. \cite{rodrigues2022tracesim}, we evaluate dedupT using ROC-AUC against baseline methods. The task is framed as a binary classification problem, where the model predicts whether a crash report is unique or a duplicate. Reports with known duplicates are labeled positive (1), while those without are labeled negative (0). The ranking score of the top result serves as the prediction score, and ROC-AUC measures the model’s ability to distinguish unique from duplicate reports across decision thresholds.

Table \ref{tab:auc-comparison} presents a comparison of various methods across all datasets. dedupT achieves the highest ROC-AUC among sequence alignment and information retrieval methods, outperforming the closest baseline, Tracesim, by over 20\% on the Netbeans dataset. It also performs competitively with deep learning methods, except on Gnome, where DeepCrash achieved the highest score. However, ROC-AUC should be interpreted alongside RR@1 in this scenario. A high ROC-AUC but low RR@1 indicates that while the model differentiates well, it fails to rank the correct duplicate first, reducing its practical utility since developers rely on top-ranked suggestions for triaging. For instance, DeepCrash achieves a high ROC-AUC on the Gnome dataset, yet its RR@1 remains low at 0.365 (Table \ref{tab:rr-comparison}), limiting its practical effectiveness for this dataset.
\begin{table}
    \small
    \centering
    \caption{ROC-AUC comparison of different approaches.}
    \resizebox{7cm}{!}{
    \begin{tabular}{lcccc}
        \toprule
        \multirow{2}{*}{\textbf{Approach}} & \multicolumn{4}{c}{\textbf{ROC-AUC}} \\
        \cmidrule(lr){2-5}
        & \textbf{Netbeans} & \textbf{Eclipse} & \textbf{Gnome} & \textbf{Ubuntu} \\
        \midrule
        Tracesim & 0.770 & 0.778 & 0.718 & 0.773 \\
        Rebucket & 0.719 & 0.777 & 0.668 & 0.671\\
        PrefixMatching & 0.755 & 0.742 & 0.796 & 0.722\\
        Needleman & 0.709 & 0.731 & 0.813 & 0.653\\
        Durfex & 0.716 & 0.761 &  -- & -- \\
        Moroo & 0.741 & 0.791 & 0.646 & 0.712\\
        Brodie & 0.627 & 0.784 & 0.740 & 0.661\\
        TF-IDF & 0.617 & 0.667 & 0.588 & 0.624\\
        S3M & 0.810 & 0.813 & 0.881 & 0.897 \\
        DeepCrash & 0.908 & 0.880 & \textbf{0.928} & 0.873 \\
        \midrule
        \textbf{dedupT} & \textbf{0.933} & \textbf{0.891} & 0.905 & \textbf{0.905} \\
        \bottomrule
    \end{tabular}
    }
    \label{tab:auc-comparison}
\end{table}

\mybox{\textbf{RQ2:} \small{dedupT achieves strong ROC-AUC while maintaining high RR@1, often outperforming IR baselines by over 20\% and providing more reliable top-ranked suggestions.}}

\subsection{\textbf{RQ3: Embedding Model Adaptation}}
\noindent\textbf{Significance of Adaptation.} Section \ref{sec:motivation} emphasized the importance of adapting or fine-tuning embedding models specifically on stack trace data. Table \ref{tab:embedding-eval} quantifies this impact by comparing Pearson and Spearman cosine correlation, as well as Euclidean similarity, before and after fine-tuning across multiple datasets. The observed improvements in Pearson and Spearman correlations indicate that fine-tuned embeddings capture semantic similarities more effectively. Additionally, the increase in Euclidean similarity suggests that embedding distances better reflect the structural relationships among stack traces.

Among the datasets, Ubuntu exhibited the lowest correlation and similarity scores before fine-tuning due to its distinct coroutine naming conventions (e.g., \texttt{IA\_\_gtk\_dialog\_run}), which differ significantly from the natural language patterns that embedding models are typically trained on. This variation makes it challenging for the model to learn meaningful representations, leading to weaker similarity measures. However, after fine-tuning, Ubuntu shows the most substantial improvement, demonstrating that domain-specific adaptation enables the model to better capture both structural and semantic relationships, even for C++ stack traces with highly specialized naming conventions.
\begin{table}
    \small
    \caption{Embedding quality evaluation before and after adapting to stacktraces. Higher correlation and similarity values indicate improved separability between duplicate and non-duplicate stack-trace pairs.}
    \centering
    \resizebox{7.5cm}{!}{
    \begin{tabular}{lccc|ccc}
        \toprule
        \multirow{2}{*}{\textbf{Dataset}} & \multicolumn{3}{c|}{\textbf{Before}} & \multicolumn{3}{c}{\textbf{After}} \\
        \cmidrule(lr){2-4} \cmidrule(lr){5-7} 
        & \textbf{Pear.} & \textbf{Spear.} & \textbf{Euclid.} & \textbf{Pear.} & \textbf{Spear.} & \textbf{Euclid.} \\
        \midrule
        Netbeans & 0.779 & 0.779 & 0.721 & 0.897 & 0.844 & 0.841 \\
        Eclipse  & 0.763 & 0.780 & 0.747 & 0.869 & 0.815 & 0.831 \\
        Gnome    & 0.738 & 0.783 & 0.750 & 0.880 & 0.837 & 0.821 \\
        Ubuntu & 0.531 & 0.477 & 0.532 & 0.781 & 0.792 & 0.705 \\
        \bottomrule
    \end{tabular}
    }
    \label{tab:embedding-eval}
\end{table}

\begin{table}
\begin{center}
\caption{Comparison of embedding model adoption methods.}
\label{table:fine-tuning-effectiveness}
\resizebox{8.5cm}{!}{
\begin{tabular}{lcccccc}
\toprule
\textbf{Dataset} & \textbf{Method} & \textbf{MRR} & \textbf{MRR Improv.} & \textbf{RR@1} & \textbf{RR@5} & \textbf{RR@10} \\
\midrule
\multirow{2}{1.5cm}{\centering \textbf{Netbeans}} & No Tuning & 0.687 & - & 0.597 & 0.790 & 0.843 \\
& Fine Tuning & \textbf{0.771} & $\uparrow\textbf{12.22}\%$ & \textbf{0.681} & \textbf{0.878} & \textbf{0.918} \\
\midrule
\multirow{2}{1.5cm}{\centering \textbf{Eclipse}} & No Tuning & 0.720 & - & 0.645 & 0.812 & 0.848 \\
& Fine Tuning & \textbf{0.791} & $\uparrow\textbf{9.86}\%$ & \textbf{0.720} & \textbf{0.880} & \textbf{0.902}  \\
\bottomrule
\end{tabular}
}
\end{center}
\vspace{-1em}
\end{table}

Beyond embedding quality, fine-tuning of the embedding model also boosts downstream retrieval performance when the duplicate classifier is trained with the fine-tuned embedding model. Table \ref{table:fine-tuning-effectiveness} shows that exclusive fine-tuning significantly improves retrieval metrics. For instance, in the Netbeans dataset, fine-tuning improves MRR by 12.22\% and RR@1 by 14\% compared to no fine-tuning (i.e., using the model off the shelf). Without fine-tuning, performance lies around the similar level to the baselines, making it an essential step for achieving meaningful improvements.

\noindent\textbf{Comparison with SOTA Embedding Models.}
We observed improvements in MRR and RR@k after fine-tuning embedding models. A natural question is how these gains compare with state-of-the-art embeddings such as OpenAI’s text-embedding-3-small \cite{openai-textemb3} or code-focused models like CodeBERT \cite{feng2020codebert}. To evaluate this, we substituted our encoder with these models and trained the classifier on their embeddings. Table \ref{table:comparison-with-sota} summarizes the results and shows our fine-tuned embedding model significantly outperforms both. Despite their strengths, these models have notable limitations. OpenAI’s embeddings are general-purpose and proprietary, cannot be fine-tuned, and incur usage-based costs, preventing domain-specific adaptation to stack traces and introducing inference-time expenses. Similarly, Code-oriented model like CodeBERT is trained on code and natural-language text but is not optimized for measuring semantic similarity across stack-trace sequences.

\begin{table}
\begin{center}
\caption{Comparison with SOTA embedding models.}
\label{table:comparison-with-sota}
\resizebox{8.5cm}{!}{
\begin{tabular}{lcccccc}
\toprule
\textbf{Dataset} & \textbf{Embedding Model} & \textbf{Vector Size} & \textbf{MRR} & \textbf{RR@1} & \textbf{RR@5} \\
\midrule
\multirow{3}{1.5cm}{\centering \textbf{Netbeans}} & text-embedding-3-small & 1536 & 0.696 & 0.603 & 0.800\\
& CodeBERT & 768 & 0.687 & 0.598 & 0.794 \\
& Ours & 768 & \textbf{0.771} & \textbf{0.681} & \textbf{0.878}  \\
\midrule
\multirow{3}{1.5cm}{\centering \textbf{Ubuntu}} & text-embedding-3-small & 1536 & 0.707 & 0.637 & 0.797 \\
& CodeBERT & 768 & 0.608 & 0.554 & 0.652 \\
& Ours & 768 & \textbf{0.786} & \textbf{0.744} & \textbf{0.838} \\
\bottomrule
\end{tabular}
}
\end{center}
\vspace{-2em}
\end{table}

To address these issues, we adopt an open-source, fine-tunable, cost-free sentence transformer model that is fully self-contained at inference. Crucially, it supports contrastive training, enabling embeddings that better distinguish duplicate crash reports by pulling semantically similar sequences closer and pushing dissimilar ones apart. This makes it a more practical and effective backbone for our approach.

\mybox{\textbf{RQ3:} \small{Fine-tuning on stack trace data improves both embedding similarity of the embedding model and retrieval performance (e.g., 12.22\% MRR improvement on Netbeans). The model also outperforms SOTA proprietary models like text-embedding-3-small.}}
\subsection{\textbf{RQ4: Ablation Study}}
\label{sec:ablation}
In this section, we evaluate a few critical design choices we made for this study.

\begin{figure}[htbp]
    \centering
    \includegraphics[width=0.8\linewidth]{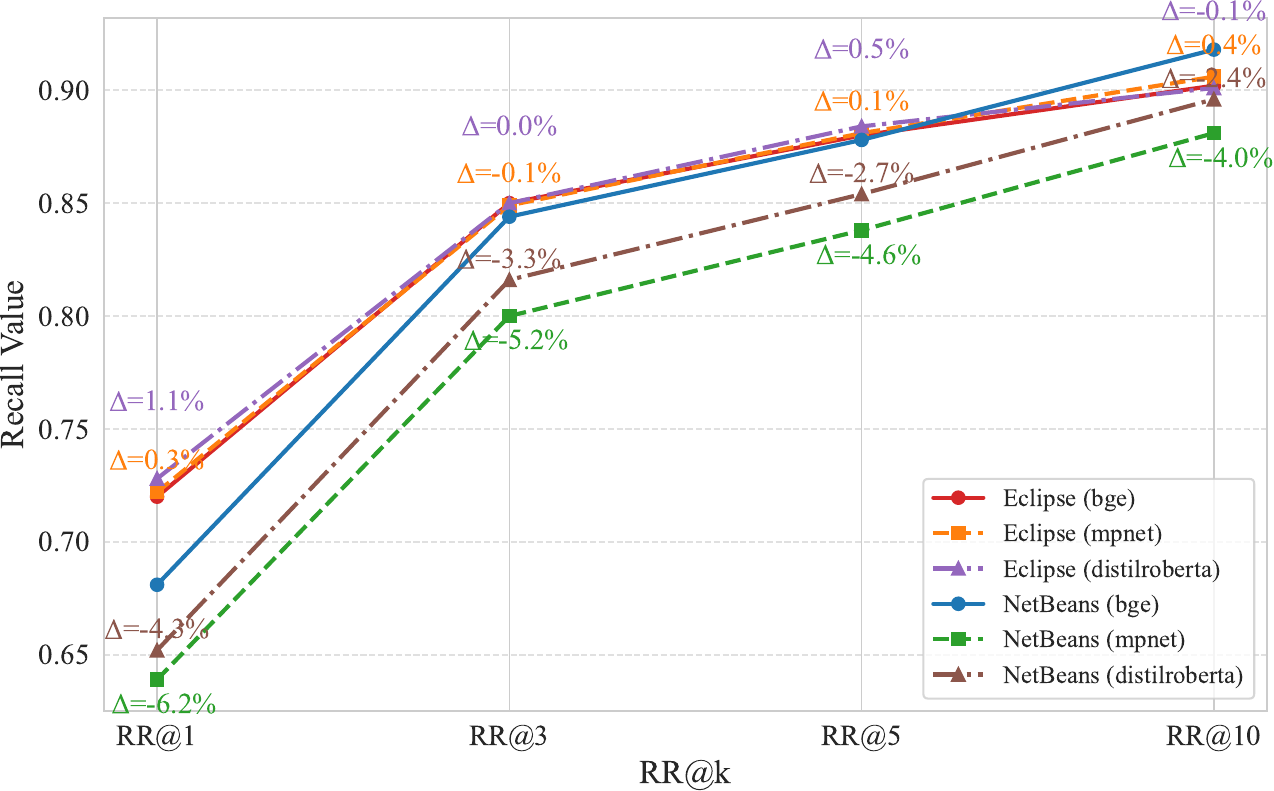}
    \caption{RR@k for fine-tuned models across datasets.}
    \label{fig:rr-vs-embeddingmodel}
    \vspace{-1em}
\end{figure}

\noindent\textbf{Pretrained Embedding Models.}
We evaluated multiple SBERT models to assess their performance. Following Section \ref{sec:approach}, all models were fine-tuned under identical settings, and a classifier was trained using their embeddings. Figure \ref{fig:rr-vs-embeddingmodel} shows RR@k and performance differences ($\Delta$) between bge and mpnet/distilroberta on Netbeans and Eclipse. A positive $\Delta$ indicates improvement over bge, while a negative value suggests the opposite.

Post fine-tuning, the performance differences across models are minimal ($ \approx 1\%$) on the Eclipse dataset. However, on the NetBeans dataset, the difference is slightly higher ($ \approx 5\%$) between bge and mpnet. This variation could arise from the differing maximum sequence lengths supported by these models and the specific nature of the datasets. The Netbeans dataset requires more tokens as we did not apply trimming to reduce the sequence length. Notably, both bge and distilroberta support a maximum sequence length of 512 tokens, while mpnet supports only 384 tokens, which contributed to the higher performance gap in this dataset. This suggests that a higher sequence length can potentially improve performance, especially in cases with longer inputs.

Overall, the results indicate that fine-tuning plays a far more significant role than the choice of the pretrained model itself. A well-optimized fine-tuning strategy can effectively mitigate model-specific differences, making multiple models viable for the task. Nonetheless, we selected bge for our experiments due to its slightly more stable performance across both datasets, ensuring consistency and reliability in our results.

\noindent\textbf{Multi-Stack Aggregation.}  
We evaluated four different stack trace aggregation techniques within our approach. Given two crash reports, \( A \) and \( B \), each containing multiple stack traces, we define \( S_A = \{s_1, s_2, \dots, s_n\} \) and \( S_B = \{q_1, q_2, \dots, q_m\} \) be the sets of stack traces for crash reports \( A \) and \( B \), respectively. We explore the following aggregation methods:

\textbf{\textit{Max Pooling.}} This method selects the most similar stack trace pair across both reports by the cosine similarity of embeddings. 
\begin{equation}
h = \max_{i, j} \text{sim}(s_i, q_j)
\end{equation} 

\textbf{\textit{Mean Pooling.}} When a crash report contains multiple stack traces, we compute the mean of their embeddings to obtain a single representative embedding:  

\begin{equation}  
h = \frac{1}{n} \sum_{i=1}^{n} e(s_i)  
\end{equation}  

where \( e(s_i) \) represents the embedding of the \( i \)-th stack trace. This aggregated embedding is then used for similarity computation.

\textit{\textbf{Multi-Head Attention Aggregation.}} We apply a multi-head attention mechanism over all stack traces to learn weighted representations dynamically:
\begin{equation}
h = \sum_{i=1}^{n} \alpha_i e_i, \quad \text{where } \alpha_i = \text{softmax}(\mathbf{q}^\top e_i)
\end{equation}
Here, \( e_i \) are the stack trace embeddings, and \( \mathbf{q} \) is a learnable query vector.  

\textbf{\textit{Parametric Max-Mean Aggregation.}} This method balances by weighting the most similar stack trace embeddings and the mean of the all embeddings. We compute this by using Equation \ref{eq:max-sim} and Equation \ref{eq:parametric-max-mean}.

Table \ref{tab:multi-stack-agg-comparison} shows that parametric max-mean and mean pooling performed the best, while multihead attention yielded the lowest performance on the Netbeans dataset. The poor performance of max pooling suggests that relying only on the most similar stack trace pair between crash reports leads to information loss. In contrast, mean pooling preserves more context by averaging all stack traces, making it more effective. Multihead attention likely struggled due to irrelevant attention weights or challenges in learning meaningful dependencies across multiple stack traces.
\begin{table}
    \small
    \caption{Comparison of multi-stack aggregation methods.}
    \centering
    \resizebox{6.5cm}{!}{
    \begin{tabular}{lcc|cc}
        \toprule
        \multirow{2}{*}{\textbf{Approach}} & \multicolumn{2}{c|}{\textbf{Netbeans}} & \multicolumn{2}{c}{\textbf{Eclipse}} \\
        \cmidrule(lr){2-3} \cmidrule(lr){4-5} 
        & \textbf{MRR} & \textbf{RR@1} & \textbf{MRR} & \textbf{RR@1} \\
        \midrule
        Max Pooling & 0.722 & 0.633 & 0.750 & 0.669 \\
        Mean Pooling &  0.761 & 0.673 & 0.786 & \textbf{0.727}  \\
        Multi-Head Attention & 0.654 & 0.566 & 0.778 & 0.707 \\
        Parametric Max-Mean & \textbf{0.771} & \textbf{0.681} & \textbf{0.791} & 0.720 \\
        \bottomrule
    \end{tabular}
    }
    \label{tab:multi-stack-agg-comparison}
    \vspace{-1em}
\end{table}


\noindent\textbf{Stack Frame Selection.} Since we encode entire stack traces at once for contextual embeddings, token limits of the PLM constrain how many frames can be processed. Exceeding this limit may result in an incomplete frame at the end because of truncation. Thus, selecting the most relevant frames, as discussed in Section \ref{sec:stack-preprocessor}, is crucial—especially for verbose Java stack traces. In contrast, shorter C/C++ subroutines allow processing more frames. Prior research found that fix-relevant methods typically occur within the top ten frames \cite{schroter2010dostacktraces}. Hence, we retain at least 10 frames and gradually increase the count to evaluate embedding similarity metrics before the adaptation stage, using this analysis to screen candidate frame budgets before downstream adaptation and classifier training.

\begin{figure}
    \centering
    \includegraphics[width=0.7\linewidth]{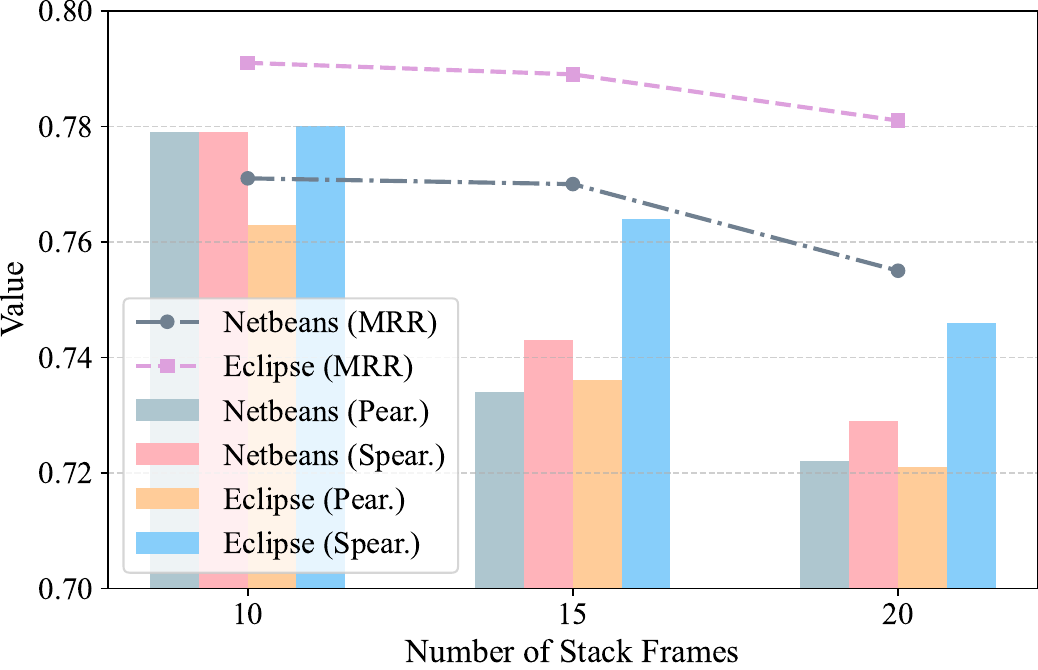}
    \caption{Pre-adaptation embedding similarity and downstream MRR for different numbers of stack frames.}
    \label{fig:frames-vs-embeddings}
    \vspace{-1em}
\end{figure}

As shown in Figure \ref{fig:frames-vs-embeddings}, increasing the number of frames generally lowers Pearson and Spearman cosine similarity scores, as additional frames alter embeddings to accommodate new information in Java stack traces. The figure also illustrates the impact of using more frames for PLM fine-tuning and finding duplicates, showing that excessive frames (20 frames) lead to lower MRR. However, extreme reduction (e.g., keeping only the exception frame) may maximize these similarity metrics but will not necessarily improve deduplication. For example, two instances of \texttt{java.lang.NullPointerException} may appear highly similar yet lack sufficient context for root cause analysis or to find an appropriate duplicate. This is especially important in contrastive learning. If single frames are sampled from duplicate and non-duplicate stack trace pairs, the same frame may appear in both positive and negative pairs. Without sufficient surrounding context, such cases can mislead the model.

\revtwo{As a practical guideline, we recommend starting with the top ten frames. Practitioners can then evaluate each candidate $N$ on a small validation set before adaptation and select the value that maximizes the Pearson and Spearman correlations between embedding cosine similarity and duplicate labels. This screening avoids retraining the full pipeline for every candidate value; the selected budget can then be validated once using downstream MRR. We advise against reducing $N$ towards a single frame, which removes the context required for contrastive learning, and $N$ should remain within the encoder's token budget (e.g., 512 tokens).}

\noindent\textbf{Stack Frame Truncation.}
We evaluate three levels of stack frame trimming within dedupT for Java stack traces: (1) No trimming (\( L_0 \)), (2) Method-level trimming (\( L_1 \)), and (3) Class-level trimming (\( L_2 \)). The trimming process removes parts of the stack frame to analyze its impact on model performance. For instance, given a stack frame \textit{org\allowbreak.example\allowbreak.package\allowbreak.Class\allowbreak.method}, \( L_0 \) retains the full frame, \( L_1 \) removes the method name \textit{org\allowbreak.example\allowbreak.package\allowbreak.Class}, and \( L_2 \) removes both the method and class name \textit{org\allowbreak.example\allowbreak.package}.  

Similar to prior studies \cite{Khvorov2021S3MSS}, we observe that transformer models also exhibit sensitivity to stack frame truncation and the effect varies across datasets. As shown in Table \ref{table:trimming-effectiveness}, increasing the trimming level negatively impacts performance on the Netbeans dataset, where the best results are achieved without trimming. In contrast, the Eclipse dataset benefits from more aggressive truncation, achieving the highest performance at class-level trimming (\( L_2 \)). This highlights that the optimal truncation level is dataset-dependent, and selecting the right level of stack frame trimming is also important for transformer-based approaches.

\begin{table}
\begin{center}
\caption{Model performance at different trimming levels.}
\label{table:trimming-effectiveness}
\resizebox{8cm}{!}{
\begin{tabular}{llcccc}
\toprule
\textbf{Dataset} & \textbf{Trimming Level} & \textbf{MRR} & \textbf{RR@1} & \textbf{RR@5} & \textbf{RR@10} \\
\midrule
\multirow{3}{1.5cm}{\centering \textbf{Netbeans}} & $L_0$ (No trim) & \textbf{0.771} & \textbf{0.681} & \textbf{0.878} & 0.918  \\
& $L_1$ (Method level) & 0.763 & 0.669 & 0.878 & \textbf{0.923}\\
& $L_2$ (Class level) & 0.726 & 0.618 & 0.851 & 0.904\\
\midrule
\multirow{3}{1.5cm}{\centering \textbf{Eclipse}} & $L_0$ (No trim) &  0.684 & 0.616 & 0.759 & 0.791\\
& $L_1$ (Method level) & 0.737 & 0.660 & 0.833 & 0.868 \\
& $L_2$ (Class level) & \textbf{0.791} & \textbf{0.720} & \textbf{0.880} & \textbf{0.902} \\
\bottomrule
\end{tabular}
}
\end{center}
\vspace{-1em}
\end{table}

\mybox{\textbf{RQ4:} \small{Fine-tuned embedding models perform similarly, with Parametric Max-Mean working better to aggregate multiple stack traces. Retaining the top 10 frames yields best results, while stack frame trimming has varying effects across datasets and should be evaluated for individual Java datasets.}}

\section{Discussion}
\label{sec:discussion}


\subsection{\textbf{\revtwo{Support for Manual Duplicate Triage}}}
\label{sec:manual-evaluation}
\revtwo{To evaluate how well dedupT's rankings support the manual confirmation of a duplicate, we compare it with TraceSim in a controlled study on two manual outcomes: (1) how often the ranked buckets support a correct confirmation and (2) the effort required to reach a decision. We aim to determine whether higher ranking positions lead to less manual inspection before a developer can confirm a duplicate.}

\noindent\textbf{\revtwo{Study Design and Protocol.}} \revtwo{We randomly sampled 100 queries from the Eclipse test set and evaluated the ranked buckets produced by dedupT and TraceSim~\cite{rodrigues2022tracesim}, the strongest baseline on Eclipse, for the same queries. A purpose-built tool (provided in the replication package) displayed each query beside one ranked candidate at a time, revealing buckets in rank order up to rank ten. A sample of this size lets us estimate ranking effectiveness on the Eclipse test set at a 95\% confidence level, and evaluating both methods on the identical queries makes the comparison paired.}

\revtwo{Within each bucket, reports were ordered chronologically from oldest to newest, placing the master report first. We did not reorder reports by duplication score, as placing each method's highest-scoring report first could bias the annotator’s decision. The annotator examined reports in this order and stopped once the visible evidence supported a decision. Reports in the same bucket that visibly repeated the same exception and call chain as an already examined member did not require separate comparison against the query crash report. The tool recorded the decision, elapsed time, number of buckets opened, and number of reports inspected.} \revtwo{The first author evaluated both rankings by comparing the exception type and the contiguous top-ten-frame call chain of each report with the query. We used this fixed evidence window because prior research found that fix-relevant methods typically occur within the top ten frames \cite{schroter2010dostacktraces}. The same evidence and stopping criteria were applied to both methods. To limit expectation and order effects, the tool concealed method identity and alternated which ranking was inspected first.}

\noindent\textbf{\revtwo{Comparative Actionability.}} \revtwo{Under the same manual protocol, the annotator accepted a candidate bucket for 93 of the dedupT queries and 84 of the TraceSim queries. Of these decisions, 89 for dedupT and 77 for TraceSim matched the ground truth. Thus, dedupT produced 12 more manually confirmed ground-truth buckets than TraceSim over the same queries. The remaining accepted buckets may reflect either annotator error or plausible related failures not grouped together in the ground truth, as discussed in Section~\ref{sec:threats}. To explain where this difference arises, we group the queries below by the predicted rank of the ground-truth bucket. These rank groups are determined from the model outputs and ground truth, whereas the examples characterize the evidence observed during manual inspection.}

\revtwo{\textit{\textbf{Rank 1:}} The ground-truth bucket is ranked first for 75 of the 100 queries by dedupT, compared with 59 by TraceSim.} \rev{For example, \revtwo{Report~\#444552 and its rank-1 prediction, Report~\#443906,} both throw \textit{ResourceException: Resource '...' does not exist} with a frame-for-frame identical top-10 frame call chain through Resource.checkExists $\rightarrow$ \textit{JavaUILabelProvider.decorateImage}, with only the runtime file path differing.} \revtwo{Both methods rank this pair first. The methods return different rank-1 buckets for 39 queries; dedupT ranks the ground-truth bucket first for 23 of these and TraceSim for 7, a significant difference under an exact McNemar test ($p = 0.005$). For instance, in Report~\#446976, dedupT ranks its ground-truth bucket (Report~\#446777) first, which has an identical chain through \textit{EditorInputPropertyTester.test}, whereas TraceSim ranks Report~\#436863, an \textit{AssertionFailedException} raised while closing a compare editor, and does not retrieve the ground-truth bucket. In this case, TraceSim's position- and frequency-based frame weights allow shared framework frames to outweigh the shorter fault-specific prefix.}

\revtwo{\textit{\textbf{Ranks 2--10:}} The ground-truth bucket appears in this range for a further 16 dedupT queries and 24 TraceSim queries, at mean ranks of 3.3 and 3.8, respectively. For Report~\#462822, the ground-truth bucket (Report~\#462483) has the same \textit{NullPointerException} at \textit{SvnNgMergeDriver.normalizeMergeSources} and an otherwise identical merge call chain; dedupT ranks it second and TraceSim ninth. TraceSim instead ranks first a different \textit{NullPointerException} from the same component but a distinct call chain. The annotator therefore reaches the same bucket after two bucket inspections with dedupT and nine with TraceSim.}

\revtwo{\textit{\textbf{Beyond rank 10:}} The ground-truth bucket is absent from the top ten for 9 dedupT queries and 17 TraceSim queries. Nevertheless, dedupT's first result shares the query's exception type in 8 of its 9 misses and its top frame in 6, compared with 10 and 5 of TraceSim's 17 misses. For instance, for Report~\#462156, dedupT retrieves a report with the same \textit{RuntimeException}, top frame, and subsequent injector chain, whereas TraceSim retrieves an unrelated \textit{JavaModelException}. Thus, even when dedupT misses the ground-truth bucket, its top-ranked results often remain associated with the failure mode being investigated.}

\revtwo{Aggregating these rank groups, dedupT attains an MRR of 0.811 on the sampled queries (95\% CI: 0.742--0.880), compared with 0.672 for TraceSim (95\% CI: 0.589--0.754). Since both methods were evaluated on the same queries, we also compare their per-query reciprocal ranks directly: the mean paired difference is 0.139 in dedupT's favor (95\% CI: 0.061--0.217), which excludes zero. Hence, 100 queries suffice to bound the ranking effectiveness of either method within $\pm$0.07--0.08 at a 95\% confidence level and to establish that dedupT's advantage is not an artifact of the sampled subset.}

\begin{table}[t]
    \small
    \centering
    \caption{Manual evaluation on 100 Eclipse queries, grouped by ground-truth bucket rank. \textbf{AB}, \textbf{AR}, and \textbf{AT} are the average buckets inspected, conservative report count, and time (seconds) per query. AR charges all members of rejected buckets and reports up to confirmation in the accepted bucket.}%
    \label{tab:ranking-validity}
    \begin{tabular}{llcccc}
        \toprule
        \textbf{Bucket Rank} & \textbf{Method} & \textbf{Cases} & \textbf{AB} & \textbf{AR} & \textbf{AT (s)} \\
        \midrule
        \multirow{2}{*}{Rank-1} & dedupT & 75 & 1.0 & \textbf{1.4} & \textbf{46} \\
         & TraceSim & 59 & 1.0 & 1.7 & 51 \\
        \midrule
        \multirow{2}{*}{Rank 2--5} & dedupT & 14 & \textbf{2.5} & \textbf{4.4} & \textbf{81} \\
         & TraceSim & 20 & 3.0 & 7.7 & 114 \\
        \midrule
        \multirow{2}{*}{Rank 6--10} & dedupT & 2 & 7.5 & 22.0 & 213 \\
         & TraceSim & 4 & \textbf{7.2} & \textbf{10.2} & \textbf{201} \\
        \midrule
        \multirow{2}{*}{Not in Top-10} & dedupT & 9 & \textbf{8.1} & \textbf{13.1} & \textbf{238} \\
         & TraceSim & 17 & 9.3 & 19.0 & 248 \\
        \midrule
        \multirow{2}{*}{\textbf{Overall}} & dedupT & 100 & \textbf{2.0} & \textbf{3.2} & \textbf{72} \\
         & TraceSim & 100 & 3.0 & 6.2 & 103 \\
        \midrule
        \multicolumn{6}{l}{\footnotesize\itshape With dedupT, it took 120 minutes across 198 bucket inspections.} \\
        \multicolumn{6}{l}{\footnotesize\itshape With TraceSim, it took 172 minutes across 305 bucket inspections.} \\
        \bottomrule
    \end{tabular}
    \vspace{-1em}
\end{table}

\noindent\textbf{\revtwo{Manual Inspection Effort.}}
\revtwo{As Table~\ref{tab:ranking-validity} shows, inspecting the dedupT rankings required approximately 120 minutes in total, compared with 172 minutes for TraceSim (a 30\% reduction). dedupT was faster for 68 of the 100 queries and required 198 bucket inspections, compared with 305 for TraceSim. Thus, the effort reduction primarily follows from placing the ground-truth bucket earlier: the annotator reached a decision after 2.0 buckets and 72 seconds per query with dedupT, on average, versus 3.0 buckets and 103 seconds with TraceSim.}

\revtwo{\textit{\textbf{Effect of bucket size.}}} \revtwo{The two methods exposed the annotator to similarly sized buckets: their rank-1 buckets contained 7.5 reports on average for dedupT and 7.0 for TraceSim, while their top-five buckets contained a total of 18.2 and 21.1 reports, respectively (median 13 for both). These averages are skewed by a thin tail, as the median rank-1 bucket holds 2 reports for both methods and 41 and 44 of their rank-1 buckets are singletons. The largest is the recurring NullPointerException in \textit{EditorInputPropertyTester.test} bucket (Report~\#446777), which had grown to 100 reports by the last query that retrieved it, yet still required reading one report to confirm. Moreover, bucket size did not translate directly into inspection depth. The annotator reached a confirmation at the first (master or the oldest) report in 79 of 93 accepted buckets for dedupT and 66 of 84 for TraceSim; the corresponding mean inspection depths were 1.4 and 1.7 reports. For dedupT, the correlation between bucket size and the number of reports inspected was only 0.14. Even accepted buckets containing more than ten reports (25.7 on average) required inspection of only 1.96 reports on average. This result is conservative with respect to model assistance because the chronological ordering provides no score-based shortcut. It also reflects the redundancy within large buckets: once several reports visibly exhibited the same exception and call chain, the annotator could decide without comparing every near-identical member.}

\revtwo{For a conservative comparison, Table~\ref{tab:ranking-validity} charges every report in a rejected bucket as inspected such that its report counts are therefore upper bounds. Under this convention, dedupT required 325 report inspections compared with 615 for TraceSim across 100 cases. The principal practical benefit is thus not smaller buckets, but fewer buckets opened and fewer reports examined before reaching a defensible decision.}

\subsection{Comparison with LLM-based Approaches}
\rev{The advent of large language models (LLMs) such as GPT-4 has greatly advanced natural language processing and is being used in many software engineering tasks \cite{zhu2023large, he2025llm}. This raises a natural question: \textit{can modern off-the-shelf LLM-based retrieval pipelines replace task-specific embedding adaptation for crash report deduplication?} To investigate this practical concern, we evaluate a representative LLM-based setup commonly used in Retrieval-Augmented Generation (RAG) \cite{lewis2020retrieval} systems.}

In this pipeline, all stack traces are embedded using OpenAI’s \texttt{text-embedding-3-small} (TE3S) model and stored in a vector database (ChromaDB \cite{chromadb}). For a given query stack trace, the top-20 most similar reports are retrieved from the existing buckets based on embedding similarity. The candidates are subsequently reranked using GPT-4o (v. 2024-08-06) in a single inference pass, where the model reasons over exception types, method/class names, call structures, etc., to produce a final ranking. Our results (Table~\ref{tab:llm-comparison}) show that TE3S with vector database alone (when no adaptation or classifier training is performed) performs poorly across both Ubuntu and NetBeans datasets, with near-zero RR@1. Reranking with GPT-4o improves MRR and RR@5 but still falls short of our stack-trace–optimized embedding model. 

\rev{We note that this evaluation differs from the comparison in Table~\ref{table:comparison-with-sota}. In Table~\ref{table:comparison-with-sota}, models such as CodeBERT and TE3S are evaluated within the same supervised learning framework used by dedupT, where a classifier is trained on top of CodeBERT embeddings. In contrast, the LLM-based pipeline here follows a retrieval-plus-reranking paradigm without task-specific fine-tuning. Thus, the comparison reflects two distinct approaches: supervised domain-adapted learning versus zero-shot LLM reranking over a retrieved subset.}

Importantly, the reranking stage is inherently limited by the initial retrieval step—if the actual duplicate crash report (i.e., the report belonging to the correct duplicate bucket) is not included among the top-$k$ retrieved candidates, the LLM cannot identify the correct bucket. Additionally, the LLM-based pipeline also suffers from high latency compared to ours ($\sim$30s vs. $\sim$2s per query) and limited scalability due to context length and cost. In contrast, our method directly optimizes embeddings for stack traces, enabling more accurate and efficient retrieval. Although LLMs can complement this approach, for example, in metadata extraction or selective reranking, specialized embeddings remain superior for the core task of similar crash report retrieval. A promising direction for future work is hybrid pipelines that combine stack-trace–aware embeddings with lightweight LLMs. Our method could first narrow the search space with high accuracy, and an LLM could then perform reranking or contextual reasoning on the top candidates.
\begin{table}[t]
\small
\centering
\caption{Comparison of our approach with LLM-based pipelines across Ubuntu and NetBeans datasets. \textbf{VDB}: Vector Database; \textbf{FT}: Fine-tuned}
\label{tab:llm-comparison}
\resizebox{\linewidth}{!}{
\begin{tabular}{llllccc}
\toprule
\textbf{Dataset} & \textbf{Sample Size} & \textbf{Approach} & \textbf{Embedder} & \textbf{MRR} & \textbf{RR@1} & \textbf{RR@5} \\
\midrule
\multirow{3}{*}{Ubuntu} 
& \multirow{3}{*}{Full test set} & VDB & TE3S & 0.118 & 0.000 & 0.103 \\
&               & VDB + GPT-4o & TE3S & 0.584 & 0.427 & 0.761 \\
&               & Ours & bge-base (FT) & \textbf{0.786} & \textbf{0.744} & \textbf{0.838} \\
\midrule
\multirow{3}{*}{NetBeans} 
& \multirow{3}{*}{200 samples} & VDB & TE3S & 0.121 & 0.000 & 0.101 \\
&             & VDB + GPT-4o & TE3S & 0.586 & 0.422 & 0.825 \\
&             & Ours & bge-base (FT) & \textbf{0.752} & \textbf{0.651} & \textbf{0.863} \\
\bottomrule
\end{tabular}}
\end{table}

\subsection{Retrieval Time}
While this study primarily focuses on accuracy, we also evaluate retrieval efficiency across deep learning methods. Table \ref{tab:retrieval-time} shows that S3M achieves the highest throughput, while DeepCrash performs the worst due to LSTM’s sequential processing overhead. Although transformers are significantly larger models, they achieve lower retrieval times by processing all tokens in parallel. However, retrieval time increases for the NetBeans and Eclipse datasets due to multiple stack trace aggregations, which introduce sequential dependencies and additional computation overhead. Since retrieval efficiency was not the primary objective, we did not optimize the pipeline in this study. Future work can enhance performance using batch processing, optimized indexing or approximate nearest neighbor search \cite{andoni2018approximatenearestneighborsearch} (e.g., FAISS \cite{douze2024faiss}) for faster retrieval.

\begin{table}
    \small
    \centering
    \caption{Comparison of retrieval time (in seconds) across datasets.}
    \resizebox{6.5cm}{!}{
    \begin{tabular}{lcccc}
        \toprule
        \textbf{Approach} & \textbf{Netbeans} & \textbf{Eclipse} & \textbf{Gnome} & \textbf{Ubuntu} \\
        \midrule
        S3M & 1.20s & 1.20s & 1.15s & 0.10s\\
        DeepCrash & 3.5s  & 3.5s & 3.25s & 1.72s \\
        Ours & 2.50s & 2.70s & 1.20s & 0.20s \\
        \bottomrule
    \end{tabular}
    }
    \label{tab:retrieval-time}
    \vspace{-1em}
\end{table}

\subsection{Applicability to Other Programming Languages}
\rev{Our evaluation focuses on Java and C/C++ due to the availability of large, well-curated datasets with labeled duplicate crash reports in these ecosystems. However, dedupT is not inherently restricted to statically typed languages. The approach operates on stack traces as structured execution sequences and does not rely on type information. The core components—contextual embeddings, contrastive adaptation, and supervised ranking—are language-agnostic and can be applied to stack traces from other languages. Adapting dedupT to additional languages would primarily require adjustments in the preprocessing stage. Different runtime environments generate stack traces with distinct formatting conventions. For example, in dynamically typed languages (e.g., Python or JavaScript), stack traces may contain shorter call chains or dynamically generated functions, and greater variability due to runtime behaviors. These characteristics may increase lexical variability across duplicates and require language-specific normalization strategies. Furthermore, as with any supervised embedding approach, fine-tuning on representative labeled data from the target ecosystem would be necessary to align the embedding space with its stack-trace distribution.}


\section{Threats to Validity} 
\label{sec:threats}
\textbf{External validity} concerns the generalizability of our findings beyond the datasets and environments used in this study. Our evaluation focuses on stack trace-based crash deduplication using specific datasets, which may not fully represent all software projects, particularly those with different logging mechanisms or programming languages. Threats to \textbf{internal validity} concerns about biases in data cleaning or filtering. Our results depend on factors such as preprocessing steps and training stability. Since DeepCrash \cite{deepcrash} had no publicly available code, we had to reproduce it based on the original paper and adapt it to our retrieval ranking pipeline, which may introduce slight variations from the original implementation. To mitigate these risks, we followed the method descriptions closely and ensured consistency in our evaluation. However, variations in data splits and potential biases in labeled duplicates may impact reproducibility. \revtwo{The study to estimate manual effort focuses on decisions based on exception types and visible call-chain similarity within the top ten frames. In practice, developers may need to inspect deeper frames, source code, or issue history, or consult other developers before confirming a duplicate. The recorded times therefore quantify focused stack-trace inspection in our tool rather than end-to-end industrial triage. All assessments were conducted by the first author, which may introduce subjectivity into the confirmation decisions and recorded inspection times. Concealing method identity and alternating the inspection order mitigated, but did not eliminate, this potential bias. Although most buckets in the studied datasets are singletons, the retrieved candidates included multi-report buckets of varying sizes. Observations may also differ across datasets and development settings. Evaluating both methods using the same queries and protocol supports a controlled within-annotator comparison; however, the absolute times and their generalizability to developer populations remain limited.}

\section{Conclusions}
\label{sec:conclusions}
We introduce dedupT, a transformer-based crash report deduplication approach that relies solely on stack traces. dedupT surpasses existing baselines in ranking duplicate crash buckets while maintaining strong performance in identifying unique reports. By fine-tuning pretrained language models on stack trace data, our method effectively captures structural and semantic similarities, improving retrieval accuracy. Future work will focus on adding support for more programming languages, enhancing efficiency for faster retrieval, etc. Additionally, we will explore reranking strategies with large language model (LLM) integration to further refine duplicate ranking performance. \rev{Furthermore, we plan to conduct controlled studies with developers in real-world crash-triage workflows to validate the observed reductions in inspection effort and examine developer behavior and the information practitioners rely on when confirming duplicates.} Overall, dedupT demonstrates the potential of treating stack traces holistically rather than as isolated frames, paving the way for more effective crash deduplication in large-scale systems.

\balance
\bibliographystyle{IEEEtran}
\bibliography{references}

@inproceedings{campbell2016unreasonable,
author = {Campbell, Joshua Charles and Santos, Eddie Antonio and Hindle, Abram},
title = {The unreasonable effectiveness of traditional information retrieval in crash report deduplication},
year = {2016},
isbn = {9781450341868},
publisher = {Association for Computing Machinery},
address = {New York, NY, USA},
doi = {10.1145/2901739.2901766},
booktitle = {Proceedings of the 13th International Conference on Mining Software Repositories},
pages = {269–280},
numpages = {12},
location = {Austin, Texas},
series = {MSR '16}
}

@article{rodrigues2022tracesim,
  title={TraceSim: An alignment method for computing stack trace similarity},
  author={Rodrigues, Irving Muller and Khvorov, Aleksandr and Aloise, Daniel and Vasiliev, Roman and Koznov, Dmitrij and Fernandes, Eraldo Rezende and Chernishev, George and Luciv, Dmitry and Povarov, Nikita},
  journal={Empirical Software Engineering},
  volume={27},
  number={2},
  pages={53},
  year={2022},
  publisher={Springer}
}

@article{Khvorov2021S3MSS,
  title={S3M: Siamese Stack (Trace) Similarity Measure},
  author={Aleksandr Khvorov and Roman Vasiliev and George A. Chernishev and Irving Muller Rodrigues and Dmitrij V. Koznov and Nikita Povarov},
  journal={2021 IEEE/ACM 18th International Conference on Mining Software Repositories (MSR)},
  year={2021},
  pages={266-270},
}

@inproceedings{Brodie2005MLMMWCS05,
  title={Quickly finding known software problems via automated symptom matching},
  author={Brodie, Mark and Ma, Sheng and Lohman, Guy and Mignet, Laurent and Modani, Natwar and Wilding, Mark and Champlin, Jon and Sohn, Peter},
  booktitle={Second International Conference on Autonomic Computing (ICAC'05)},
  pages={101--110},
  year={2005},
  organization={IEEE}
}

@article{Sabor2017DURFEXAF,
  title={DURFEX: A Feature Extraction Technique for Efficient Detection of Duplicate Bug Reports},
  author={Korosh Koochekian Sabor and Abdelwahab Hamou-Lhadj and Alf Larsson},
  journal={2017 IEEE International Conference on Software Quality, Reliability and Security (QRS)},
  year={2017},
  pages={240-250},
}

@inproceedings{Fader14wikihow,
    author    = {Anthony Fader and Luke Zettlemoyer and Oren Etzioni},
    title     = {{Open Question Answering Over Curated and Extracted
                Knowledge Bases}},
    booktitle = {KDD},
    year      = {2014}
}

@misc{bajaj2018msmarcohumangenerated,
      title={MS MARCO: A Human Generated MAchine Reading COmprehension Dataset}, 
      author={Payal Bajaj and Daniel Campos and Nick Craswell and Li Deng and Jianfeng Gao and Xiaodong Liu and Rangan Majumder and Andrew McNamara and Bhaskar Mitra and Tri Nguyen and Mir Rosenberg and Xia Song and Alina Stoica and Saurabh Tiwary and Tong Wang},
      year={2018},
      eprint={1611.09268},
      archivePrefix={arXiv},
      primaryClass={cs.CL},
}

@inproceedings{Homma2017DetectingDQ,
  title={Detecting duplicate questions with deep learning},
  author={Homma, Yushi and Sy, Stuart and Yeh, Christopher},
  booktitle={Proceedings of the International Conference on Neural Information Processing Systems (NIPS)},
  pages={25964--25975},
  year={2016}
}

@INPROCEEDINGS{Modani2007,
  author={Modani, Natwar and Gupta, Rajeev and Lohman, Guy and Syeda-Mahmood, Tanveer and Mignet, Laurent},
  booktitle={2007 IEEE 23rd International Conference on Data Engineering Workshop}, 
  title={Automatically Identifying Known Software Problems}, 
  year={2007},
  volume={},
  number={},
  pages={433-441},
  doi={10.1109/ICDEW.2007.4401026}}

@inproceedings{Bartz2008FindingSF,
  title={Finding similar failures using callstack similarity},
  author={Bartz, Kevin and Stokes, Jack W and Platt, John and Kivett, Ryan and Grant, David and Calinoiu, Silviu and Loihile, Gretchen},
  booktitle={SysML08: Third Workshop on Tackling Computer Systems Problems with Machine Learning Techniques},
  year={2008}
}

@INPROCEEDINGS{Dhaliwal2011,
  author={Dhaliwal, Tejinder and Khomh, Foutse and Zou, Ying},
  booktitle={2011 27th IEEE International Conference on Software Maintenance (ICSM)}, 
  title={Classifying field crash reports for fixing bugs: A case study of Mozilla Firefox}, 
  year={2011},
  volume={},
  number={},
  pages={333-342},
  doi={10.1109/ICSM.2011.6080800}}

@INPROCEEDINGS{Lerch2013,
  author={Lerch, Johannes and Mezini, Mira},
  booktitle={2013 17th European Conference on Software Maintenance and Reengineering}, 
  title={Finding Duplicates of Your Yet Unwritten Bug Report}, 
  year={2013},
  volume={},
  number={},
  pages={69-78},
  doi={10.1109/CSMR.2013.17}}

@inproceedings{wong2014boosting,
  title={Boosting bug-report-oriented fault localization with segmentation and stack-trace analysis},
  author={Wong, Chu-Pan and Xiong, Yingfei and Zhang, Hongyu and Hao, Dan and Zhang, Lu and Mei, Hong},
  booktitle={2014 IEEE international conference on software maintenance and evolution},
  pages={181--190},
  year={2014},
  organization={IEEE}
}

@INPROCEEDINGS{schroter2010dostacktraces,
  author={Schroter, Adrian and Schröter, Adrian and Bettenburg, Nicolas and Premraj, Rahul},
  booktitle={2010 7th IEEE Working Conference on Mining Software Repositories (MSR 2010)}, 
  title={Do stack traces help developers fix bugs?}, 
  year={2010},
  volume={},
  number={},
  pages={118-121},
  doi={10.1109/MSR.2010.5463280}}

@inproceedings{devlin-etal-2019-bert,
    title = "{BERT}: Pre-training of Deep Bidirectional Transformers for Language Understanding",
    author = "Devlin, Jacob  and
      Chang, Ming-Wei  and
      Lee, Kenton  and
      Toutanova, Kristina",
    editor = "Burstein, Jill  and
      Doran, Christy  and
      Solorio, Thamar",
    booktitle = "Proceedings of the 2019 Conference of the North {A}merican Chapter of the Association for Computational Linguistics: Human Language Technologies, Volume 1 (Long and Short Papers)",
    month = jun,
    year = "2019",
    address = "Minneapolis, Minnesota",
    publisher = "Association for Computational Linguistics",
    doi = "10.18653/v1/N19-1423",
    pages = "4171--4186"
}

@inproceedings{Reimers2019SentenceBERTSE,
  title={Sentence-BERT: Sentence Embeddings using Siamese BERT-Networks},
  author={Nils Reimers and Iryna Gurevych},
  booktitle={Conference on Empirical Methods in Natural Language Processing},
  year={2019}
}

@article{matthew2018elmo,
  author       = {Matthew E. Peters and
                  Mark Neumann and
                  Mohit Iyyer and
                  Matt Gardner and
                  Christopher Clark and
                  Kenton Lee and
                  Luke Zettlemoyer},
  title        = {Deep contextualized word representations},
  journal      = {CoRR},
  volume       = {abs/1802.05365},
  year         = {2018},
  eprinttype    = {arXiv},
  eprint       = {1802.05365}
}

@article{roberta2019,
  author       = {Yinhan Liu and
                  Myle Ott and
                  Naman Goyal and
                  Jingfei Du and
                  Mandar Joshi and
                  Danqi Chen and
                  Omer Levy and
                  Mike Lewis and
                  Luke Zettlemoyer and
                  Veselin Stoyanov},
  title        = {RoBERTa: {A} Robustly Optimized {BERT} Pretraining Approach},
  journal      = {CoRR},
  volume       = {abs/1907.11692},
  year         = {2019},
  eprinttype    = {arXiv},
  eprint       = {1907.11692}
}

@article{vaswani2017attention,
  title={Attention is all you need},
  author={Vaswani, Ashish and Shazeer, Noam and Parmar, Niki and Uszkoreit, Jakob and Jones, Llion and Gomez, Aidan N and Kaiser, {\L}ukasz and Polosukhin, Illia},
  journal={Advances in neural information processing systems},
  volume={30},
  year={2017}
}

@ARTICLE{rakha2018,
  author={Rakha, Mohamed Sami and Bezemer, Cor-Paul and Hassan, Ahmed E.},
  journal={IEEE Transactions on Software Engineering}, 
  title={Revisiting the Performance Evaluation of Automated Approaches for the Retrieval of Duplicate Issue Reports}, 
  year={2018},
  volume={44},
  number={12},
  pages={1245-1268},
  doi={10.1109/TSE.2017.2755005}}

@inproceedings{deepcrash,
author = {Chao, Liu and Qiaoluan, Xie and Yong, Li and Yang, Xu and Hyun-Deok, Choi},
title = {DeepCrash: deep metric learning for crash bucketing based on stack trace},
year = {2022},
isbn = {9781450394567},
publisher = {Association for Computing Machinery},
address = {New York, NY, USA},
doi = {10.1145/3549034.3561179},
booktitle = {Proceedings of the 6th International Workshop on Machine Learning Techniques for Software Quality Evaluation},
pages = {29–34},
numpages = {6},
location = {Singapore, Singapore},
series = {MaLTeSQuE 2022}
}

@article{chen2020contrastive,
  author       = {Ting Chen and
                  Simon Kornblith and
                  Mohammad Norouzi and
                  Geoffrey E. Hinton},
  title        = {A Simple Framework for Contrastive Learning of Visual Representations},
  journal      = {CoRR},
  volume       = {abs/2002.05709},
  year         = {2020},
  eprinttype    = {arXiv},
  eprint       = {2002.05709}
}

@article{asudani2023impact,
author = {Asudani, Deepak Suresh and Nagwani, Naresh Kumar and Singh, Pradeep},
title = {Impact of word embedding models on text analytics in deep learning environment: a review},
year = {2023},
issue_date = {Sep 2023},
publisher = {Kluwer Academic Publishers},
address = {USA},
volume = {56},
number = {9},
issn = {0269-2821},
doi = {10.1007/s10462-023-10419-1},
journal = {Artif. Intell. Rev.},
month = feb,
pages = {10345–10425},
numpages = {81}
}

@inproceedings{xiao2024cpack,
  title={C-pack: Packed resources for general chinese embeddings},
  author={Xiao, Shitao and Liu, Zheng and Zhang, Peitian and Muennighoff, Niklas and Lian, Defu and Nie, Jian-Yun},
  booktitle={Proceedings of the 47th international ACM SIGIR conference on research and development in information retrieval},
  pages={641--649},
  year={2024}
}

@misc{henderson2017efficient,
    title={Efficient Natural Language Response Suggestion for Smart Reply},
    author={Matthew Henderson and Rami Al-Rfou and Brian Strope and Yun-hsuan Sung and Laszlo Lukacs and Ruiqi Guo and Sanjiv Kumar and Balint Miklos and Ray Kurzweil},
    year={2017},
    eprint={1705.00652},
    archivePrefix={arXiv},
    primaryClass={cs.CL}
}

@inproceedings{burges2005learning,
  title={Learning to rank using gradient descent},
  author={Burges, Chris and Shaked, Tal and Renshaw, Erin and Lazier, Ari and Deeds, Matt and Hamilton, Nicole and Hullender, Greg},
  booktitle={Proceedings of the 22nd international conference on Machine learning},
  pages={89--96},
  year={2005}
}

@article{Kingma2014AdamAM,
  title={Adam: A method for stochastic optimization},
  author={Kingma, Diederik P and Ba, Jimmy},
  journal={arXiv preprint arXiv:1412.6980},
  year={2014}
}

@article{srivastava2014dropout,
  title={Dropout: a simple way to prevent neural networks from overfitting},
  author={Srivastava, Nitish and Hinton, Geoffrey and Krizhevsky, Alex and Sutskever, Ilya and Salakhutdinov, Ruslan},
  journal={The journal of machine learning research},
  volume={15},
  number={1},
  pages={1929--1958},
  year={2014},
  publisher={JMLR. org}
}

@inproceedings{vasiliev2020tracesim,
author = {Vasiliev, Roman and Koznov, Dmitrij and Chernishev, George and Khvorov, Aleksandr and Luciv, Dmitry and Povarov, Nikita},
title = {TraceSim: a method for calculating stack trace similarity},
year = {2020},
isbn = {9781450381246},
publisher = {Association for Computing Machinery},
address = {New York, NY, USA},
doi = {10.1145/3416505.3423561},
booktitle = {Proceedings of the 4th ACM SIGSOFT International Workshop on Machine-Learning Techniques for Software-Quality Evaluation},
pages = {25–30},
numpages = {6},
location = {Virtual, USA},
series = {MaLTeSQuE 2020}
}

@inproceedings{dang2012rebucket,
  title={Rebucket: A method for clustering duplicate crash reports based on call stack similarity},
  author={Dang, Yingnong and Wu, Rongxin and Zhang, Hongyu and Zhang, Dongmei and Nobel, Peter},
  booktitle={2012 34th International Conference on Software Engineering (ICSE)},
  pages={1084--1093},
  year={2012},
  organization={IEEE}
}

@inproceedings{moroo2017reranking,
  title={Reranking-based Crash Report Deduplication.},
  author={Moroo, Akira and Aizawa, Akiko and Hamamoto, Takayuki},
  booktitle={SEKE},
  volume={17},
  pages={507--510},
  year={2017}
}

@inproceedings{rodrigues2022fast,
  title={FaST: A linear time stack trace alignment heuristic for crash report deduplication},
  author={Rodrigues, Irving Muller and Aloise, Daniel and Fernandes, Eraldo Rezende},
  booktitle={Proceedings of the 19th International Conference on Mining Software Repositories},
  pages={549--560},
  year={2022}
}

@article{shi2022abaci,
  title={Abaci-finder: Linux kernel crash classification through stack trace similarity learning},
  author={Shi, Heyuan and Wang, Guyu and Fu, Ying and Hu, Chao and Song, Houbing and Dong, Jian and Tang, Kun and Liang, Kai},
  journal={Journal of Parallel and Distributed Computing},
  volume={168},
  pages={70--79},
  year={2022},
  publisher={Elsevier}
}

@inproceedings{yang2025kermit,
  title={KERMIT: A BERT-Based Classification Method for Linux Kernel Crashes Through Stack Trace},
  author={Yang, Yunshan and Dong, Pan and Jiang, Renshuang and Fang, Xiaoxiang and Yu, Qirui and Li, Bao and Zhang, Jianfeng and Ding, Yan},
  booktitle={International Conference on Intelligent Computing},
  pages={263--274},
  year={2025},
  organization={Springer}
}

@inproceedings{kudo2018sentencepiece,
  title={SentencePiece: A simple and language independent subword tokenizer and detokenizer for neural text processing},
  author={Kudo, Taku and Richardson, John},
  booktitle={Proceedings of the 2018 conference on empirical methods in natural language processing: System demonstrations},
  pages={66--71},
  year={2018}
}

@inproceedings{andoni2018approximatenearestneighborsearch,
  title={Approximate nearest neighbor search in high dimensions},
  author={Andoni, Alexandr and Indyk, Piotr and Razenshteyn, Ilya},
  booktitle={Proceedings of the International Congress of Mathematicians: Rio de Janeiro 2018},
  pages={3287--3318},
  year={2018},
  organization={World Scientific}
}

@article{douze2024faiss,
  title={The faiss library},
  author={Douze, Matthijs and Guzhva, Alexandr and Deng, Chengqi and Johnson, Jeff and Szilvasy, Gergely and Mazar{\'e}, Pierre-Emmanuel and Lomeli, Maria and Hosseini, Lucas and J{\'e}gou, Herv{\'e}},
  journal={IEEE Transactions on Big Data},
  year={2025},
  publisher={IEEE}
}

@inproceedings{shibaev2024stack,
  title={Stack trace deduplication: Faster, more accurately, and in more realistic scenarios},
  author={Shibaev, Egor and Sushentsev, Denis and Golubev, Yaroslav and Khvorov, Aleksandr},
  booktitle={2025 IEEE International Conference on Software Analysis, Evolution and Reengineering (SANER)},
  pages={511--521},
  year={2025},
  organization={IEEE}
}

@inproceedings{he2020duplicate,
  title={Duplicate bug report detection using dual-channel convolutional neural networks},
  author={He, Jianjun and Xu, Ling and Yan, Meng and Xia, Xin and Lei, Yan},
  booktitle={Proceedings of the 28th International Conference on Program Comprehension},
  pages={117--127},
  year={2020}
}

@inproceedings{xiao2020hindbr,
  title={Hindbr: Heterogeneous information network based duplicate bug report prediction},
  author={Xiao, Guanping and Du, Xiaoting and Sui, Yulei and Yue, Tao},
  booktitle={2020 IEEE 31st international symposium on software reliability engineering (ISSRE)},
  pages={195--206},
  year={2020},
  organization={IEEE}
}

@inproceedings{rodrigues2020soft,
  title={A soft alignment model for bug deduplication},
  author={Rodrigues, Irving Muller and Aloise, Daniel and Fernandes, Eraldo Rezende and Dagenais, Michel},
  booktitle={Proceedings of the 17th International Conference on Mining Software Repositories},
  pages={43--53},
  year={2020}
}

@INPROCEEDINGS{sun2011rep,
  author={Sun, Chengnian and Lo, David and Khoo, Siau-Cheng and Jiang, Jing},
  booktitle={2011 26th IEEE/ACM International Conference on Automated Software Engineering (ASE 2011)}, 
  title={Towards more accurate retrieval of duplicate bug reports}, 
  year={2011},
  volume={},
  number={},
  pages={253-262},
  doi={10.1109/ASE.2011.6100061}}

@inproceedings{remil2024deeplsh,
  title={Deeplsh: Deep locality-sensitive hash learning for fast and efficient near-duplicate crash report detection},
  author={Remil, Youcef and Bendimerad, Anes and Mathonat, Romain and Raissi, Chedy and Kaytoue, Mehdi},
  booktitle={Proceedings of the IEEE/ACM 46th International Conference on Software Engineering},
  pages={1--12},
  year={2024}
}

@misc{huggingface-all-mpnet,
  author       = {HuggingFace},
  title        = {all-mpnet-base-v2},
  url          = {https://huggingface.co/sentence-transformers/all-mpnet-base-v2},
  lastaccessed = {March 13, 2025},
}

@misc{chromadb,
  author       = {ChromaDB},
  title        = {ChromaDB},
  url          = {https://www.trychroma.com/},
  lastaccessed = {March 13, 2025},
}

@misc{openai-textemb3,
  author       = {OpenAI},
  title        = {text-embedding-3-small},
  url          = {https://platform.openai.com/docs/models/text-embedding-3-small},
  lastaccessed = {August 20, 2025},
}

@misc{huggingface-distilroberta,
  author       = {HuggingFace},
  title        = {distilroberta-v1},
  url          = {https://huggingface.co/sentence-transformers/all-distilroberta-v1},
  lastaccessed = {March 13, 2025},
}

@inproceedings{karita2019comparative,
  title={A comparative study on transformer vs rnn in speech applications},
  author={Karita, Shigeki and Chen, Nanxin and Hayashi, Tomoki and Hori, Takaaki and Inaguma, Hirofumi and Jiang, Ziyan and Someki, Masao and Soplin, Nelson Enrique Yalta and Yamamoto, Ryuichi and Wang, Xiaofei and others},
  booktitle={2019 IEEE automatic speech recognition and understanding workshop (ASRU)},
  pages={449--456},
  year={2019},
  organization={IEEE}
}

@article{Deberta,
  title={Deberta: Decoding-enhanced bert with disentangled attention},
  author={He, Pengcheng and Liu, Xiaodong and Gao, Jianfeng and Chen, Weizhu},
  journal={arXiv preprint arXiv:2006.03654},
  year={2020}
}

@article{bashiri2024comprehensive,
  title={Comprehensive review and comparative analysis of transformer models in sentiment analysis},
  author={Bashiri, Hadis and Naderi, Hassan},
  journal={Knowledge and Information Systems},
  volume={66},
  number={12},
  pages={7305--7361},
  year={2024},
  publisher={Springer}
}

@article{Zhang2019BERTScoreET,
  title={BERTScore: Evaluating Text Generation with BERT},
  author={Tianyi Zhang and Varsha Kishore and Felix Wu and Kilian Q. Weinberger and Yoav Artzi},
  journal={ArXiv},
  year={2019},
  volume={abs/1904.09675},
}

@article{li2024pre,
  title={Pre-trained language models for text generation: A survey},
  author={Li, Junyi and Tang, Tianyi and Zhao, Wayne Xin and Nie, Jian-Yun and Wen, Ji-Rong},
  journal={ACM Computing Surveys},
  volume={56},
  number={9},
  pages={1--39},
  year={2024},
  publisher={ACM New York, NY}
}

@article{feng2020codebert,
  title={Codebert: A pre-trained model for programming and natural languages},
  author={Feng, Zhangyin and Guo, Daya and Tang, Duyu and Duan, Nan and Feng, Xiaocheng and Gong, Ming and Shou, Linjun and Qin, Bing and Liu, Ting and Jiang, Daxin and others},
  journal={arXiv preprint arXiv:2002.08155},
  year={2020}
}

@article{zhu2023large,
  title={Large language models for information retrieval: A survey},
  author={Zhu, Yutao and Yuan, Huaying and Wang, Shuting and Liu, Jiongnan and Liu, Wenhan and Deng, Chenlong and Chen, Haonan and Liu, Zheng and Dou, Zhicheng and Wen, Ji-Rong},
  journal={ACM Transactions on Information Systems},
  volume={44},
  number={1},
  pages={1--54},
  year={2025},
  publisher={ACM New York, NY}
}

@article{he2025llm,
  title={Llm-based multi-agent systems for software engineering: Literature review, vision, and the road ahead},
  author={He, Junda and Treude, Christoph and Lo, David},
  journal={ACM Transactions on Software Engineering and Methodology},
  volume={34},
  number={5},
  pages={1--30},
  year={2025},
  publisher={ACM New York, NY}
}

@article{lewis2020retrieval,
  title={Retrieval-augmented generation for knowledge-intensive nlp tasks},
  author={Lewis, Patrick and Perez, Ethan and Piktus, Aleksandra and Petroni, Fabio and Karpukhin, Vladimir and Goyal, Naman and K{\"u}ttler, Heinrich and Lewis, Mike and Yih, Wen-tau and Rockt{\"a}schel, Tim and others},
  journal={Advances in neural information processing systems},
  volume={33},
  pages={9459--9474},
  year={2020}
}

\end{document}